\documentclass[12pt]{article}
\pdfoutput=1

\usepackage{amsmath,amsfonts,amssymb,epsfig,comment,xspace,listings,cite}
\usepackage{graphicx,caption,subcaption}

\usepackage{slashed}

\usepackage{listings}

\usepackage{booktabs}

\usepackage{tabularx, multirow}

\usepackage[width=0.9\textwidth,margin={30pt,30pt},font=small,labelfont=bf]{caption}
\numberwithin{equation}{section}

\usepackage{hyperref}

\usepackage{xcolor}
\hypersetup{
    colorlinks,
    linkcolor={red!0!black},
    citecolor={blue!50!black},
    urlcolor={blue!80!black}
}

\usepackage{graphicx}
\usepackage{cancel}
\usepackage{cite}
\usepackage{color}
\usepackage{bm}
\usepackage{multirow}
\usepackage{xspace}

\def\twomat[#1,#2][#3,#4]{\left( \begin{array}{cc} #1 & #2 \\ #3 & #4 \end{array} \right)}
\def\twoa[#1,#2][#3,#4]{\left( \begin{array}{cc} #1 & #2 \\ #3 & #4 \end{array} \right)}

\def\thv[#1,#2,#3]{\left( \begin{array}{c} #1 \\ #2 \\ #3 \end{array} \right)}
\def\twv[#1,#2]{\left( \begin{array}{c} #1 \\ #2 \end{array} \right)}

\def\GeV{\ensuremath{\mathrm{GeV}}\xspace}

\def\mueff{\mu_{\rm eff}}
\def\Bmueff{B_{\mu,\, \mathrm{eff}}}

\def\SARAH{{\tt SARAH}\xspace}

\def\HiggsBounds{{\tt HiggsBounds5}\xspace}

\def\ov{\overline}
\def\bra{\langle}
\def\ket{\rangle}

\def\MS{M_S}

\def\MS{M_{{\rm\scriptscriptstyle SUSY}}}

\def\s2b{s_{2\beta}}
\def\c2b{c_{2\beta}}

\def\ov{\overline}

\def\GeV{\ \mathrm{GeV}}

\def\beq{\begin{equation}}
\def\eeq{\end{equation}}
\def\bal{\begin{align}}
\def\eal{\end{align}}

\newcommand{\exclude}[1]{}
\def\nn{\nonumber}

\def\nn{\nonumber}

\begin{document}

\begin{flushright}
\end{flushright}
\begin{center}

\vspace{1cm}
{\LARGE{\bf  W boson mass in minimal Dirac gaugino scenarios}}

\vspace{1cm}

\large{  Karim Benakli$^a$ \let\thefootnote\relax\footnote{$^a$kbenakli@lpthe.jussieu.fr},
Mark Goodsell$^b$\footnote{$^b$goodsell@lpthe.jussieu.fr},
Wenqi Ke$^c$ \footnote{$^c$wke@lpthe.jussieu.fr} and
Pietro Slavich$^d$ \footnote{$^d$slavich@lpthe.jussieu.fr} 
}

\vspace{0.5 cm}

{
\emph{Sorbonne Universit\'e, CNRS, Laboratoire de Physique Th\'eorique et Hautes Energies (LPTHE), F-75005 Paris, France. 
}}


\end{center}
\vspace{0.7cm}

\abstract{We investigate the conditions for alignment in Dirac Gaugino models with minimal matter content. This leads to several scenarios, including an aligned Dirac Gaugino NMSSM that allows a light singlet scalar. We then investigate the compatibility of minimal Dirac Gaugino models with an enhanced W boson mass, using a new precise computation of the quantum corrections included in the code SARAH 4.15.0.}

\newpage

\tableofcontents

\setcounter{footnote}{0}

\section{Introduction}
\label{Sec:introduction}



While constraints on heavy Higgs bosons in supersymmetric models are rather stringent at large $\tan \beta$, excluding masses above a TeV, at small to moderate $\tan \beta$ direct searches do not place significant limits; only indirect constraints from $B\rightarrow s \gamma$ limit a heavy charged Higgs boson to be above $568$ GeV, roughly independent of $\tan \beta$ \cite{Misiak:2017bgg}. On the other hand, in the Minimal Supersymmetric Standard Model (MSSM), this region is likely excluded for an additional neutral Higgs boson below a few hundred GeV due to modifications to the SM-like Higgs boson couplings. This has led to a lot of interest in extensions of the MSSM (or variants of the Two Higgs Doublet Model) where \emph{alignment without decoupling} is possible \cite{Davidson:2005cw,BhupalDev:2014bir,Bernon:2015qea,Bernon:2015wef,Carena:2013ooa,Carena:2015moc,Haber:2017erd,Coyle:2019exn}, that is where the mixing between the SM-like Higgs boson and other scalars is minimised so that it aligns with the expectation values and has SM-like couplings.

Dirac Gaugino models 
\cite{Fayet:1978qc,Polchinski:1982an,Hall:1990hq,Fox:2002bu,Nelson:2002ca,Antoniadis:2006uj,Belanger:2009wf,Benakli:2008pg,Benakli:2011kz}, (see also, for example,
\cite{Kribs:2007ac,Amigo:2008rc,Benakli:2009mk,Choi:2009ue,Benakli:2010gi,Choi:2010gc,%
Carpenter:2010as,Abel:2011dc,Davies:2011mp,Benakli:2011vb,Frugiuele:2011mh,%
Itoyama:2011zi,Bertuzzo:2012su,Davies:2012vu,Frugiuele:2012pe,%
Frugiuele:2012kp,Benakli:2012cy,Dudas:2013gga,Benakli:2013msa,Benakli:2015ioa,Arvanitaki:2013yja,Itoyama:2013sn,Chakraborty:2013gea,Csaki:2013fla,Itoyama:2013vxa,Beauchesne:2014pra,%
Benakli:2014daa,%
Bertuzzo:2014bwa,Diessner:2014ksa,Benakli:2014cia,Goodsell:2014dia,Busbridge:2014sha,Chakraborty:2014sda,Ding:2015wma,Alves:2015kia,Alves:2015bba,Carpenter:2015mna,Martin:2015eca,Goodsell:2015ura,Kotlarski:2016zhv,Benakli:2016ybe,Braathen:2016mmb,Chakraborty:2017dfg,Chalons:2018gez,Carpenter:2020hyz,Goodsell:2020lpx,Carpenter:2020evo}) accommodate scenarios where alignment without decoupling  is \emph{automatic} at tree-level\cite{Antoniadis:2006uj,Benakli:2018vqz,Benakli:2018vjk,Benakli:2018ldd,Benakli:2019yaq,Ellis:2016gxa}. Under the assumption of an $N=2$ supersymmetry in the gauge sector at some scale, these models contain an approximate $SU(2)_R$ R-symmetry which guarantees the tree-level alignment \cite{Benakli:2018vjk}. An investigation of the effects of quantum corrections showed that it is even radiatively stable \cite{Benakli:2018vqz}, with competing effects partially cancelling.

These models have many interesting phenomenological properties, and have been extensively studied in the literature. They involve, at a minimum, an extension of the MSSM by three adjoint chiral superfields, one for each gauge group; the fermions from these pair with the gauginos to give them a Dirac mass. This means the presence of new scalar fields, in singlet, triplet and octet representations.

Actually, it has not adequately been investigated to what extent the \emph{singlet} could be light in such models. One condition for this to be the case is that it should not disturb the couplings of the light Higgs -- in other words, we should have some amount of alignments without decoupling. ATLAS and CMS both give constraints on the overall signal strength of the Higgs boson $\mu$ to be \cite{CMS:2020gsy,ATLAS:2020qdt}:
\begin{align}
\mu =& 1.06 \pm 0.07 \ (\mathrm{ATLAS}), \qquad \mu = 1.02^{+0.07}_{-0.06} \ (\mathrm{CMS}).
\end{align}
If the Higgs boson $h$ mixes with an inert singlet $s$, then we can write the mass eigenstates $\tilde{h}, S$ as 
\begin{align}
\twv[h,s] =& \twoa[S_{11}, S_{12}][-S_{12},S_{11}]\twv[\tilde{h},S]
\end{align}
then we will find that 
\begin{align}
\mu =& |S_{11}|^2 \le 1.
\end{align}
Hence if we allow a $3\sigma$ deviation from the ATLAS result, we require 
\begin{align}
1-|S_{11}|^2 = |S_{12}|^2 \le 0.15 \longrightarrow |S_{12}| < 0.39.
\end{align}
While this still allows a moderate amount of mixing, the larger the mixing between the flavour eigenstates, the stronger the direct search bounds on the singlet will be. Therefore in this work we will consider the conditions for an approximate alignment in which the light Higgs mixes neither with the Heavy Higgs nor with the singlet (the triplet being decoupled). This will lead to a scenario that we refer to as the \emph{aligned DGNMSSM}.

Recently, the CDF experiment reported a new measurement of the mass of the W boson \cite{CDF:2022hxs}. Compared to the SM prediction \cite{Awramik:2003rn,Degrassi:2014sxa,CDF:2022hxs}, this gives as averages (combined Tevatron+LEP\cite{OPAL:2005rdt,L3:2005fft,ALEPH:2006cdc,DELPHI:2008avl,ALEPH:2013dgf,D0:2012kms,CDF:2012gpf,CDF:2013dpa,CDF:2022hxs}):
\begin{align}
M_W^{\rm Tevatron+LEP} =& 80424 \pm 9\ \mathrm{MeV}, \qquad M_W^{\rm SM} = 80356 \pm 6\ \mathrm{MeV}.
\end{align}
If we take the central value of the top quark mass to be $172.89$ GeV then the central SM prediction becomes $80352$ MeV \cite{Heinemeyer:2013dia}.
These differ by $7$ standard deviations, although measurements at the LHC \cite{ATLAS:2017rzl,LHCb:2021bjt} also differ from the combination of Tevatron+LEP by $4$ standard deviations, so at this stage confirmation is required by other experiments. Nevertheless, a modification to the W boson mass is one of the most generic effects of new light particles coupling to the electroweak sector, so such a hint is tantalising. 

It has generally been assumed in Dirac Gaugino models that the adjoint scalars should be heavy; indeed, the requirement that the triplet scalar vacuum expectation value (vev) must be very small compared to the Standard Model Higgs one -- otherwise it would generate a large $\rho$ parameter - is usually ensured by giving the triplet a heavy mass. Amusingly, following the new measurement, the simplest explanation for the enhanced W boson mass is exactly an expectation value for the neutral component of such a triplet. In this work we shall investigate that possibility in minimal Dirac Gaugino models.

Such a triplet scalar also comes along with electroweak fermions, which can modify the quantum corrections too. Therefore a \emph{precise} computation is required. While a preliminary such computation was performed for the MRSSM \cite{Athron:2022isz} using an update to {\tt FlexibleSUSY} \cite{Athron:2014wta,Athron:2017fvs}, and a related computation was performed for the same model in \cite{Diessner:2019ebm}, that model lacks a natural enhancement to the W boson quantum corrections. In this work we introduce a similarly precise computation in the package {\tt SARAH-4.15.0} and use it to examine the compatibility of our aligned DGNMSSM, along with four other scenarios -- the ``MSSM without $\mu$ term,'' the MDGSSM, the aligned MDGSSM and the general DGNMSSM -- with the new measurement of the W boson mass, or a naive world average value of $M_W^{\rm world\ average} = 80411 \pm 15$ MeV.

This work is organised as follows. In section \ref{SEC:MODELS} we summarise the essential details of the class of Dirac Gaugino models, including the vacuum minimisation conditions and mass matrices. The conditions for alignment are reviewed and a comparison is made with the cases of the MSSM and NMSSM. We also introduce the different variants we shall consider: the MSSM without $\mu$-term; the MDGSSM, the DGNMSSM and the aligned DGNMSSM. In Section \ref{SEC:MW} we will study the predictions for all of these classes of models for the W boson mass, examining in particular the effects of a precise computation of the quantum effects. We present our conclusions in section \ref{SEC:CONCLUSIONS}.

%
 
%
%

\section{Dirac Gaugino Models with Automatic Tree-level Alignment}
\label{SEC:MODELS}


%
%

\subsection{Field content and interactions}

We shall consider in this work the extension of the MSSM by a minimal
matter content to allow Dirac Gaugino masses, as in \cite{Benakli:2011kz,Benakli:2012cy}. The
additional superfields consist of three chiral multiplets, in adjoint
representations of the SM gauge group factors (DG-adjoints): a singlet
$\mathbf{S}$, an $SU(2)_W$ triplet $\mathbf{T}^a$, and an $SU(3)_C$
octet $\mathbf{O}^a$.
If we require gauge-coupling unification, even more states should be
added to the model. For instance, for an $(SU(3))^3$ Grand
Unification, the minimal set of chiral multiplets includes also extra
Higgs-like doublets $\mathbf{R}_{u,d}$ as well as two pairs of
vector-like right-handed electron $\mathbf{{E'}}_{1,2}$ in
$(\mathbf{1} ,\mathbf{1})_{1}$ and $\mathbf{\tilde{{E'}}}_{1,2}$ in $
(\mathbf{1} ,\mathbf{1})_{-1}$. We will not consider these states
here.
 
In order to develop an intuition for the different interactions
involved, it is helpful to consider a simple picture where the model
descends from a supersymmetric theory in $D$ dimensions. The different
states can appear in different sectors: some live in the whole
$D$-dimensional bulk, others are localised on four-dimensional
hyper-surfaces (branes) at points of the extra dimensions of
coordinates $x_i = \{ x^a_i\}$, $a=5, \cdots, D$. The corresponding
Lagrangian can be written as

\begin{equation}
\int d^Dx  \, \, \mathcal{L} = \int d^Dx  \, \,  \{ \mathcal{L}_{bulk} + \sum_i  \delta^{(D-4)} (x -x_i)  \, \,   \, \,  \mathcal{L}^{(i)}_{boundaries} \}~,
\end{equation}
where we have not explicitly written the metric factors. The
four-dimensional theory arises after a truncation keeping only the
compactification zero modes:

\begin{eqnarray}
\mathcal{L}^{4d} &= & \mathcal{L}^{4d}_{bulk} + \mathcal{L}_{boundaries}~,  \\
\mathcal{L}^{4d}_{bulk}  &= & \int d^{D-4}x  \, \, \mathcal{L}_{bulk} ~, \\
 \mathcal{L}_{boundaries}   &= & \int d^{D-4}x  \, \, \sum_i  \delta^{(D-4)} (x -x_i)  \, \,   \, \,  \mathcal{L}^{(i)}_{boundaries}~.
\label{boundary-bulk-Lagrangian}
\end{eqnarray}

A tree-level alignment  in the Higgs sector appears in a class of models where the bulk theory leads to a four-dimensional Lagrangian with
interactions governed by an $N=2$ extended SUSY. In
particular, the SM gauge fields and the DG-adjoint
fields arise as $N=2$ vector supermultiplets, and the two Higgs chiral
superfields $\mathbf{H}_{d}$ and $\mathbf{H}_{d}$ form an $N=2$
hypermultiplet, interacting through the superpotential
\begin{align}
\mathcal{L}^{4d}_{bulk} \supset \int d^2 \theta  \, \,   \{\mu  \, \,   \mathbf{H}_{u} \cdot \mathbf{H}_{d}+\lambda_{S}  \, \,  \mathbf{S}  \, \, \mathbf{H}_{u} \cdot \mathbf{H}_{d}+ 2 \lambda_{T}  \, \,  \mathbf{H}_{d} \cdot \mathbf{T}\mathbf{H}_{u} \}~,
\label{N=2Lagrangian}
\end{align}
where $\mathbf{T} \equiv \frac12 \sigma^a\, \mathbf{T}^a$, and the dot
product is defined as

\begin{equation}
\mathbf{H}_{u} \cdot \mathbf{H}_{d} \equiv \epsilon_{i j} \mathbf{H}_{u}^{i} \mathbf{H}_{d}^{j}=\mathbf{H}_{u}^{+} \mathbf{H}_{d}^{-} - \mathbf{H}_{u}^{0} \mathbf{H}_{d}^{0}~.
\end{equation}

The $N=2$ SUSY has a global $SU(2)_R$ R-symmetry that rotates between
the generators of the two $N=1$ supercharges. The scalar components
$S$, $T^a$ of $ \mathbf{S}$ and $\mathbf{T}^a $, respectively, are
singlets of $SU(2)_R$.  This R-symmetry rotates then between the
auxiliary fields $F_\Sigma^a$ of the adjoint superfields
$\mathbf{\Sigma}^a \in \{ \mathbf{S},\mathbf{T}^a \}$ and the
auxiliary component $D^a$ of the corresponding chiral gauge
superfields $\mathcal{W}^{a}_{i\, \alpha}$ for $U(1)_Y$ and
$SU(2)_W$. This implies that
\begin{align}
( \quad Re(F_\Sigma^a) \quad ,  \quad D^a \quad , \quad  {Im(F_\Sigma^a)} \quad )
\label{Auxil}
\end{align}
form a triplet of $SU(2)_R$. As a consequence, in order that the
interactions (\ref{N=2Lagrangian}) of $ \mathbf{S}$ and $\mathbf{T}^a$
with the two Higgs doublets preserve $SU(2)_R$, the couplings
$\lambda_S$ and $\lambda_T$ must be related through\footnote{In the discussion of the W boson mass,
we shall relax this condition and study also generic Minimal Dirac Gaugino models with arbitrary values for $\lambda_{S}$ and $\lambda_{T}$. All of the description of the models presented in this section holds for these models except for the $N=2$ SUSY and $SU(2)_R$ global R-symmetry that are broken.}
\begin{equation}
\lambda_{S}=  \frac{g_Y}{\sqrt{2}}, \quad {\rm and} \quad \lambda_{T}=  \frac{g_2}{\sqrt{2}}
\label{eq:SUSYcoupling}
\end{equation}
to the couplings $g_Y$ and $g_2$ of the $U(1)_Y$ and $SU(2)_W$ gauge
groups, respectively. Below the scale where the $N=2$ SUSY is broken
to $N=1$, these relationships are spoiled by a small amount through
renormalisation group running, so in numerical evaluations we must
treat the couplings $\lambda_S$ and $\lambda_T$ as independent
parameters.
%

In addition to the $SU(2)_R$ R-symmetry which is broken in $N=1$
(chiral) sectors, there is a global $U(1)_R$ R-symmetry under which the
superspace coordinates $\theta^\alpha$ carry a $-1$ charge.  The
$U(1)_R$ charges of the $\mathbf{H}_u $ and $\mathbf{H}_d$ superfields
are $R_{H_u}$ and $R_{H_d}$, respectively. They are arbitrary but
subject to the constraint $R_{H_u} +R_{H_d} =2$. The DG-adjoint
superfields $\mathbf{S}, \mathbf{T}^a$, and $\mathbf{O}^a$ are
R-neutral. Below, we shall classify the different $N=1$ interactions
following whether they preserve or break the $U(1)_R$ symmetry.

The boundary Lagrangian can be split into different contributions:
\begin{equation}
\mathcal{L}^{4d}_{boundaries} ~=~ \mathcal{L}^{bulk}_{localised} +  \int d^2 \theta  \, \,  \{ W_{Yukawa}  + W_{DG} + W_{NR}  \} \,+\, \Delta \mathcal{L}^{s o f t} \, .
\label{Superpotentials}
\end{equation}
Here, we denote by $\mathcal{L}^{bulk}_{localised}$ kinetic and
interaction terms already present in the bulk theory
$\mathcal{L}^{4d}_{bulk}$ but appearing with relative coefficients that violate $N=2$ supersymmetry. Such terms can a priori be present because the boundary theory preserves only $N=1$ SUSY, thus the coefficients of these terms are less constrained. 
Here, for simplicity, we assume such terms to vanish at tree
level, to be only generated by quantum loops after supersymmetry breaking, and will therefore
be accounted for in our analysis, at least in part, through the radiative corrections. 

Also in (\ref{Superpotentials}), we have the
usual MSSM Yukawa superpotential $W_{Yukawa}$ with the couplings
responsible for the quark and lepton masses:
\begin{eqnarray}
W_{Yukawa} &=& Y_u^{ij}\, \mathbf{U^c}_i \mathbf{Q}_j \cdot \mathbf{H}_u - Y_d^{ij}\, \mathbf{D^c}_i \mathbf{Q}_j\cdot \mathbf{H}_d - Y_e^{ij}\, \mathbf{E^c}_i \mathbf{L}_j \cdot \mathbf{H_d} ~,
\label{YukawaSuperpotentials}
\end{eqnarray}
which arises on the brane where the matter field supermultiplets are
localised. 

In this work, we consider a typical scale for the soft terms, for example squarks or gaugino masses,  to be in the phenomenologically interesting range $m_{soft} \sim {\rm TeV}$. If we denote by $\Lambda$ a higher scale, for instance related to supersymmetry breaking messenger mass scale or to the Planck scale, then we can consider the relative strength of the diverse SUSY-breaking terms as an expansion in powers of $\frac{m_{soft}}{\rm \Lambda}$. We will assume that SUSY-breaking terms in the gauge sector preserve the
$U(1)_R$ R-symmetry, giving rise to Dirac gaugino masses, while
Majorana masses might be generated only by higher-order interaction
terms, therefore suppressed by additional powers of hidden-sector couplings and/or $\frac{m_{soft}}{\rm \Lambda}$ where $\Lambda$ could be the Planck scale (for gravity-induced effects). The effective superpotential for the Dirac gaugino masses reads
\begin{eqnarray}
W_{DG} &=&  \sqrt{2} \theta^\alpha \bigg[ m_{DY} \mathbf{S} \mathcal{W}_{1\,\alpha} + 2m_{D2} \mathrm{tr} (\mathbf{T} \mathcal{W}_{2\,\alpha}) + 2m_{D3} \mathrm{tr} (\mathbf{O} \mathcal{W}_{3\,\alpha})    \bigg] ~,
\label{GauginomassesSuperpotentials}
\end{eqnarray}
where $\mathbf{O} \equiv \frac12 \lambda^a \mathbf{O}^a$, and
$\mathcal{W}_{i\,\alpha}$ are the chiral gauge-strength superfields.
Finally, the superpotential $W_{NR}$ contains terms that break
explicitly the $U(1)_R$ R-symmetry:
\begin{eqnarray}
W_{NR}&=&  \xi_S \,\mathbf{S}  +\frac{M_{S}}{2} \,\mathbf{S}^{2} +\frac{\kappa}{3} \,\mathbf{S}^{3}
+\lambda_{ST} \,\mathbf{S} \operatorname{tr}(\mathbf{T} \mathbf{T}) +\lambda_{SO} \,\mathbf{S} \operatorname{tr}\left(\mathbf{O} \mathbf{O}\right) 
\nn \\ 
&&+ M_{T} \operatorname{tr}(\mathbf{T} \mathbf{T}) +M_{O} \operatorname{tr}\left(\mathbf{O} \mathbf{O}\right) + \frac{\lambda_{O}}{3}\operatorname{tr}\left(\mathbf{O}\mathbf{O}\mathbf{O}\right)~.
 \label{R-violatingSuperpotentials}
\end{eqnarray}

The soft SUSY-breaking Lagrangian $\Delta \mathcal{L}^{s o f
t}$ can in turn be split in two parts. The first contains the scalar
mass and interaction terms that preserve the $U(1)_R$ R-symmetry:
\begin{eqnarray}
-\Delta \mathcal{L}^{s o f t}_R &=&m_{H_{u}}^{2}\left|H_{u}\right|^{2}+m_{H_{d}}^{2}\left|H_{d}\right|^{2}+m_{S}^{2}|S|^{2}
+2 m_{T}^{2} \operatorname{tr}\left(T^{\dagger} T\right)
+2 m_{O}^{2} \operatorname{tr}\left(O^{\dagger} O\right)
\nonumber\\[2mm]
&+&(m_{Q}^2)^{ij}Q^\dagger_iQ_j + (m_{U}^2)^{ij}U^{c\dagger}_iU^c_j
+ (m_{D}^2)^{ij}D^{c\dagger}_iD^c_j
+(m_{L}^2)^{ij}L^\dagger_iL_j + (m_{E}^2)^{ij}E^{c\dagger}_iE^c_j
\nonumber\\[2mm]
&+&\biggr( t_S S  +\frac{1}{2} B_{S} S^{2}+\frac{1}{3}  T_{\kappa} S^{3}
+ T_{ST} S \,\mathrm{tr} (TT) +T_{SO} S\, \mathrm{tr} (O O) 
\nonumber\\
&&~~~+ B_{T} \,\operatorname{tr}(T T)
+ B_{O} \operatorname{tr} \left(O O\right)+ \frac{1}{3} T_O \mathrm{tr} (OOO)
+ \text {h.c.}\biggr) ~.
\label{R-conserving-scalar}
\end{eqnarray}
The second part of $\Delta \mathcal{L}^{s o f t}$ contains the scalar
mass and interaction terms that break the $U(1)_R$ R-symmetry:
\begin{eqnarray}
-\Delta \mathcal{L}^{s o f t}_{NR}&\supset&B_{\mu} \, H_{u} \cdot H_{d}
+ T_{S} \, S H_{u} \cdot H_{d}+2 \,T_{T} \, H_{d} \cdot T H_{u}
\nonumber\\[2mm]
&+& T_u^{ij}\, {U^c}_i {Q}_j \cdot {H}_u - T_d^{ij}\, {D^c}_i {Q}_j \cdot {H}_d - T_e^{ij}\, {E^c}_i {L}_j \cdot {H_d} ~+~\text{h.c.}~,
\label{R-violating-scalar}
\end{eqnarray}
as well as the Majorana mass terms $M_i$ (with $i=1,2,3$) for the
gauginos. In general, the mechanisms that break R-symmetry and SUSY
could be independent of each other, hence in
eqs.~(\ref{R-conserving-scalar}) and (\ref{R-violating-scalar}) we
refrained from defining the soft SUSY-breaking trilinear couplings as
proportional to the corresponding superpotential couplings.
In this work we assume that the soft SUSY-breaking
Higgs-sfermion-sfermion interactions in the second line of
eq.~(\ref{R-violating-scalar}) are suppressed with respect to the
R-conserving sfermion mass terms in
eq.~(\ref{R-conserving-scalar}). This can be realised in our
$D$-dimensional picture if the quark and lepton superfields are
localised on a brane that differs from the one where the breaking of
the R-symmetry takes place.

Since our study focuses on the electroweak sector of Dirac Gaugino
models, we assume for simplicity that the scalar octet $O^a$ is heavy
and can be integrated out of the theory.  To insulate the singlet
sector from threshold corrections involving the heavy octet, we also
neglect the singlet-octet interaction term proportional to
$\lambda_{SO}$ in the R-violating part of the superpotential,
eq.~(\ref{R-violatingSuperpotentials}), as well as the analogous term
proportional to $T_{SO}$ in the R-conserving part of the soft
SUSY-breaking Lagrangian, eq.~(\ref{R-conserving-scalar}). Similarly, since they cannot appear in some of our scenarios, for simplicity in the following we shall also neglect $\lambda_{ST}, T_{ST}$ and the tadpole terms $t_S, \xi_S$.  
 
\subsection{The electroweak scalar sector and alignment}

We can now discuss the neutral scalar sector of this class of
models. The vacuum expectation values\,\footnote{It should be
emphasised that, throughout this work, we assume that CP symmetry is
not spontaneously broken by the vacuum. Therefore, all vevs are real.}
(vevs) of the neutral components of the doublets $H_d$ and $H_u$ are
related by $v_{u}^{2}+v_{d}^{2}=v^{2}$, where $v \simeq 246$ GeV is
the electroweak scale, and we define $\tan\beta = v_u/v_d$. 
The neutral singlet and triplet scalars $S$ and $T^0$ obtain vevs
$\bra S \ket = v_S/\sqrt{2}$ and $\bra T^0 \ket = v_T/\sqrt{2}$,
respectively. These lead to effective $\mu$ and $B_\mu$ parameters:
\begin{eqnarray}
  \label{eq:mueff}
  \mueff &\equiv& \mu + \frac{1}{\sqrt{2}} ( \lambda_S v_S + \lambda_T v_T), \\
  \Bmueff &\equiv& B_\mu + \frac{1}{\sqrt{2}}(\lambda_S M_S + T_S)v_S +  \frac{1}{\sqrt{2}}(\lambda_T M_T + T_T)v_T + \frac{1}{2} \lambda_S \kappa v_S^2. 
\end{eqnarray}
The vevs $v_S$ and $v_T$ are then determined as a solution for the
coupled cubic equations:
\begin{eqnarray}
\kappa^2 v_S^3+ \frac{1}{\sqrt{2}}(T_\kappa +3 \kappa M_S) v_S^2 + \tilde{m}_{SR}^2  v_S ~~~~~~~~~~~~~~~~~~~~~~~~~~~~~~~~~~~~~~~&&\nn\\
 \label{minvS}
+ \frac{v^2}{2} \bigg[ \sqrt{2} \lambda_S \mueff - g_Y m_{DY} c_{2\beta} -( \frac{1}{\sqrt{2}} T_S + \frac{1}{\sqrt{2}} \lambda_S M_S + \lambda_S v_S \kappa)s_{2\beta}\bigg] &=&0,
 \end{eqnarray}
\begin{equation}
           \label{minvT}
 \tilde{m}_{TR}^2 v_T + \frac{v^2}{2} \bigg[\sqrt{2} \lambda_T \mueff + g_2 m_{D2} c_{2\beta} - \frac{1}{\sqrt{2}}(T_T + \lambda_T M_T) s_{2\beta}   \bigg] ~=~0,
\end{equation}
where
\begin{eqnarray}
  \label{eq:mSR}
  \tilde{m}_{SR}^2 &\equiv& M_S^2 + m_S^2 + B_S + 4 m_{DY}^2, \nn\\
  \tilde{m}_{TR}^2 &\equiv& M_T^2 + m_T^2 + B_T + 4 m_{D2}^2,
\end{eqnarray}
are effective mass-squared parameters (at zero expectation value) for the real components $S_R$ and $T_R^0$ of
the neutral singlet and triplet scalars (the analogous masses for the
imaginary components $S_I$ and $T_I^0$ are $\tilde{m}_{SI}^2 = M_S^2 + m_S^2 -
B_S$ and $\tilde{m}_{TI}^2 = M_T^2 + m_T^2 - B_T$, respectively).
We know that $v_T$ must be small -- namely, less than a few GeV -- to
avoid an overlarge tree-level $\Delta \rho$, so to a good
approximation we can set $v_T=0$ in the vacuum minimisation equation
for $v_S$, eq.~(\ref{minvS}), and decouple it from the one for $v_T$,
eq.~(\ref{minvT}); this would allow the cubic equation for $v_S$ to be
solved using standard techniques. However, the current state of
technology for the computation of loop corrections assumes that we
take expectation values as being valid for the true minimum of the
full quantum-corrected potential, so in our numerical studies we must
take them as \emph{inputs}. Especially for $v_T$ this can lead to
complications; see \cite{Braathen:2021fyq} for a recent discussion of
this issue.

To discuss the alignment in the Higgs sector, it is now convenient to
introduce the so-called Higgs basis for the two doublets,
\begin{equation}
  \Phi_{1} \equiv \frac{v_{d} \Phi_{d}+v_{u} \Phi_{u}}{v},
  \quad \Phi_{2} \equiv \frac{-v_{u} \Phi_{d}+v_{d} \Phi_{u}}{v}~,
\label{eq:Higgsbasis}
\end{equation}
where we defined for convenience two doublets with positive
hypercharge, $\Phi_{u}^{j}\equiv H_{u}^{j}$ and $\Phi_{d}^{j} \equiv
- \epsilon_{i j} H_{d}^{* j}$. In the Higgs basis the two doublets can
be decomposed as
\begin{equation}
    \Phi_{1}=\left(\begin{array}{c}
G^{+} \\
(v+h+i G^0) / \sqrt{2}
\end{array}\right), \quad \Phi_{2}=\left(\begin{array}{c}
H^{+} \\
(H + i A) / \sqrt{2}
\end{array}\right) ~,
\label{eq:Higgsdecomp}
\end{equation}
i.e., only the neutral component of $\Phi_1$ has a non-zero vev, and
the would-be-Goldstone bosons, $G^{\pm}$ and $G^0$, all lie in
$\Phi_1$. In general, the neutral CP-even fields $h$ and $H$ mix with
the neutral CP-even components of the singlet and the triplet.  In the
basis $\{ h, H, S_R, T_R^0\}$, the tree-level mass matrix reads
\begin{align}
  \left(\mathcal{M}^2\right)_{\rm {\scriptscriptstyle tree}}
  =& \left( \begin{array}{cccc} M_Z^2 + \Delta_h s_{2\beta}^2 & \Delta_h s_{2\beta} c_{2\beta} & \Delta_{hs} & \Delta_{ht} \\
\Delta_h s_{2\beta} c_{2\beta} & M_A^2 - \Delta_h s_{2\beta}^2 & \Delta_{Hs} & \Delta_{Ht} \\
 \Delta_{hs}   & \Delta_{Hs} & \tilde{m}_S^2 & \lambda_S \lambda_T \frac{v^2}{2} \\
 \Delta_{ht}   & \Delta_{Ht}  & \lambda_S \lambda_T \frac{v^2}{2} & \tilde{m}_T^2 \end{array} \right), 
\end{align}
where
\begin{eqnarray}
  \label{eq:deltah}
  \Delta_h &=& \frac{v^2}{2} (\lambda_S^2 + \lambda_T^2) - M_Z^2 ,\\
    \label{eq:deltahs}
\Delta_{hs} &=& v\bigg[ \sqrt{2} \lambda_S \mueff - g_Y m_{DY} c_{2\beta} -( \frac{1}{\sqrt{2}} (T_S + \lambda_S M_S) + v_S \kappa \lambda_S) s_{2\beta} \bigg]\\
  \Delta_{Hs}&=& v \bigg[ -(\frac{1}{\sqrt{2}} (T_S + \lambda_S M_S) + v_S \kappa \lambda_S) c_{2\beta} + g_Y m_{DY} s_{2\beta}\bigg],\\
  \Delta_{ht} &=& v \bigg[ \sqrt{2} \lambda_T \mueff
    + g_2 m_{D2} c_{2\beta} - \frac{1}{\sqrt{2}}(T_T + \lambda_T M_T) s_{2\beta} \bigg], \\
  \Delta_{Ht} &=&  -v \bigg[ \frac{1}{\sqrt{2}} (T_T + \lambda_T M_T) c_{2\beta} + g_2 m_{D2} s_{2\beta} \bigg],\\
  M_A^2 &=& \frac{2\Bmueff}{s_{2\beta}},\\
  \tilde{m}_S^2 &=& \tilde{m}_{SR}^2 + \lambda_S^2 \frac{v^2}{2} - \kappa \lambda_S \frac{v^2}{2} s_{2\beta} + 3 \kappa^2 v_S^2 + \sqrt{2} v_S( T_\kappa + 3\kappa M_S) ,\\
  \tilde{m}_T^2 &=& \tilde{m}_{TR}^2 + \lambda_T^2 \frac{v^2}{2}.
\end{eqnarray}


Exact alignment in the Higgs sector is obtained when one of the
eigenstates of the CP-even mass matrix -- in this work, we take it to
be the lightest one -- is aligned in field space with the direction of
the SM Higgs vev, and thus has SM-like couplings to gauge bosons and
matter fermions.  This is equivalent to requiring that $h$ itself be
an eigenstate of ${\cal M}^2$, or in other words that ${\cal
  M}^2_{1j}=0$ with $j=2,3,4$. If we make the reasonable assumption that the triplet is heavy, then this can be relaxed to just $j=2,3$. In addition, we will also refer in this work to cases where the singlet can be light without being potentially ruled out by direct searches. In this case we will require the supplementary condition that $\mathcal{M}^2_{23} =0$.

\bigskip

We start our discussion by focusing on the alignment between the two
doublets.  The use in eq.~(\ref{eq:deltah}) of the $N=2$ condition
for the singlet and triplet superpotential couplings, see
eq.~(\ref{eq:SUSYcoupling}), implies $\Delta_h=0$ and $M_h=M_Z$,
i.e.~alignment is automatically realised at the tree level in this
class of models, and the tree-level mass of the SM-like Higgs boson is
independent of $\tan\beta$ but well below the value observed at the
LHC. It is however well known that, in SUSY models, the radiative
corrections to the Higgs mass matrix play a crucial role in lifting
the prediction for the mass of the SM-like Higgs boson up to the
observed value. Moreover, the radiative corrections to the condition
in eq.~(\ref{eq:SUSYcoupling}) for the superpotential couplings of the
adjoint superfields can become relevant if the scale $M_{N=2}$ where
the $N=2$ SUSY is broken to $N=1$ is much larger than the scale where
the Higgs mass matrix is computed. All of these corrections inevitably
affect also the condition for alignment in the Higgs sector. As was
discussed in ref.~\cite{Benakli:2018vqz}, in DG models the element
that mixes the two doublets in the loop-corrected mass matrix can be
recast as
\begin{equation}
{\cal M}^2_{12} ~=~
\frac{1}{\tan\beta}\left[{\cal M}^2_{11} - M_Z^2\right]
-\, \frac{v^2\,\tan\beta}{1+\tan^2\beta}\,
\left[\left( \lambda_S^2 - \frac{~g^{\prime\,2}}{2}\right)
+\left( \lambda_T^2 - \frac{g^2}{2}\right)\right]
+\,(...)~,
\label{eq:alignMDGSSM}
\end{equation}
where ${\cal M}^2_{11}$ contains the dominant one-loop contribution
from top and stops. The latter consists in a term enhanced by
$y_t^4\,\ln(\MS^2/m_t^2)$, where $y_t$ is the top Yukawa coupling and
$\MS$ denotes for simplicity a common soft SUSY-breaking mass
parameter for the stops. The second term in eq.~(\ref{eq:alignMDGSSM})
accounts for the deviation of the superpotential couplings from the
$SU(2)_R$ condition in eq.~(\ref{eq:SUSYcoupling}), and the ellipses
denote one-loop top/stop contributions that are suppressed by small
ratios of parameters, one-loop contributions that involve couplings
other than $y_t$, and higher-loop contributions. Close to alignment,
the loop-corrected mass-matrix element ${\cal M}^2_{11}$ can be
empirically identified with the observed mass of the SM-like Higgs
boson, $M_h^2 \approx 2\,M_Z^2$. Therefore, eq.~(\ref{eq:alignMDGSSM})
shows that the radiative corrections included in ${\cal M}^2_{11}$
tend to destroy the tree-level alignment in the Higgs sector of DG
models. However, when $M_{N=2}$ is large the evolution of $\lambda_S$
and $\lambda_T$ down to the scale where the Higgs mass matrix is
computed makes the second term in eq.~(\ref{eq:alignMDGSSM}) negative,
and partially compensates for the misalignment induced by the top/stop
contributions.

It is instructive to compare the condition for doublet alignment in DG
models with the analogous conditions in the MSSM and in the NMSSM. In
the case of the MSSM, discussed e.g.~in
refs.~\cite{Carena:2013ooa,Carena:2014nza}, one finds
\begin{equation}
{\cal M}^2_{12} ~=~
\frac{1}{\tan\beta}\left[{\cal M}^2_{11} - M_Z^2\,c_{2\beta}\right]
+\,\frac{6\,y_t^2\,m_t^2\,\mu\,A_t}{16 \pi^2\,\MS^2}\,
\left(1- \frac{A_t^2}{6\,\MS^2}\right)\,+\,(...)~,
\label{eq:alignMSSM}
\end{equation}
where $A_t \equiv T_u^{33}/y_t$ is the soft SUSY-breaking
Higgs-stop-stop interaction parameter, and again the ellipses denote
sub-dominant terms. It appears that, in the MSSM, the alignment
condition ${\cal M}^2_{12} = 0$ can be realised radiatively when a
large value of $\tan\beta$ suppresses the first term in
eq.~(\ref{eq:alignMSSM}), while the parameters $\MS$, $A_t$ and $\mu$
combine in such a way that the second term is large and negative. In
contrast, in DG models the contributions to ${\cal M}^2_{12}$
analogous to the second term in eq.~(\ref{eq:alignMSSM}) are
suppressed by the assumption that $A_t \ll \MS$, see the comments
after eq.~(\ref{R-violating-scalar}), thus doublet alignment cannot be
realised in this way. We remark however that, even with the $N=2$
condition for the superpotential couplings, in DG models ${\cal
  M}^2_{12}$ is smaller by a factor between $2$ and $3$ -- depending
on $\tan\beta$, which we assume to be greater than $1$ -- with respect
to the case of the MSSM with small $A_t$.

In the case of the NMSSM, discussed e.g.~in
refs.~\cite{Carena:2015moc, Coyle:2019exn}, the mixing between $h$ and
$H$ is given by
\begin{equation}
{\cal M}^2_{12} ~=~
\frac{1}{\tan\beta}\left[{\cal M}^2_{11} - M_Z^2\,c_{2\beta}
-\lambda_S^2 \,v^2 s^2_\beta\right]
+\,\frac{6\,y_t^2\,m_t^2\,\tilde\mu \,A_t}{16 \pi^2\,\MS^2}\,
\left(1- \frac{A_t^2}{6\,\MS^2}\right)\,+\,(...)~,
\label{eq:alignNMSSM}
\end{equation}
where $\tilde\mu \equiv \mu + \lambda_S\, v_S/\sqrt2$. Comparing with
the case of the MSSM, eq.~(\ref{eq:alignMSSM}), we see that the
condition ${\cal M}^2_{12}=0$ can be realised even in the absence of a
large contribution from the terms proportional to $\mu_{\rm eff}
A_t/\MS^2$, as long as the singlet-doublet superpotential coupling
takes values in the range $\lambda_S^2 \approx (3\!-\!4)\, M_Z^2/v^2$,
where the exact numerical coefficient depends on the value of
$\tan\beta$. As first pointed out in ref.~\cite{Carena:2015moc}, this
condition singles out the region of the NMSSM parameter space where
$\lambda_S \approx 0.7 \pm 0.05$, a much larger value than would be
implied by the $SU(2)_R$ condition in DG models.

To summarise, the $SU(2)_R$ R-symmetry implies exact alignment at the tree
level in the Higgs-doublet sector of the DG models, but the alignment
is partially spoiled by the radiative corrections that are necessary
to obtain a realistic value for the SM-like mass. Alignment in the
MSSM can be realised only through radiative corrections, for large
$\tan\beta$ and for specific choices of the parameters in the stop
sector. Finally, doublet alignment in the NMSSM can be realised even
without the help of radiative corrections for an appropriate choice of
$\lambda_S$, which -- differently from the DG case with $SU(2)_R$ R-symmetry -- is treated as a
free parameter.

\bigskip

The second condition for Higgs alignment in DG models is ${\cal
M}^2_{13} = 0$, i.e.~vanishing mixing between $h$ and $S_R$. Including
the dominant contributions from stop loops, we find: 
%
\begin{equation} 
\mathcal{M}_{13}^2 ~=~ \Delta_{hs}
- \frac{6 y_t \lambda_S c_\beta}{16\pi^2} m_t (A_t - \mu_{\rm eff}\cot\beta)
\ln\frac{\MS^2}{Q^2}\,+\,(...)~,
\label{eq:alignNMSSMs}
\end{equation}
where the tree-level mixing term $\Delta_{hs}$ is given in
eq.~(\ref{eq:deltahs}), $\mu_{\rm eff}$ is given in
eq.~(\ref{eq:mueff}), and $Q$ is the renormalisation scale at which
the parameters entering $\Delta_{hs}$ are expressed. We assumed again
a common soft SUSY-breaking mass term $\MS$ for the stops, and we
neglected terms suppressed by powers of $m_t^2/\MS^2$.
The various terms that contribute to $\Delta_{hs}$ arise from
different sectors of the $D$-dimensional picture discussed earlier in
this section: namely, $\mu$ and $\lambda_S$ enter the bulk
superpotential in eq.~(\ref{N=2Lagrangian}); $m_{1D}$ enters the
R-conserving boundary superpotential in
eq.~(\ref{GauginomassesSuperpotentials}); $M_S$ and $\kappa$ enter the
R-violating boundary superpotential in
eq.~(\ref{R-violatingSuperpotentials}); $T_S$ enters
the R-violating SUSY-breaking Lagrangian in
eq.~(\ref{R-violating-scalar}). Therefore, even if we assume the $N
=2$ SUSY relation of eq.~(\ref{eq:SUSYcoupling}) between $\lambda_S$
and $g_Y$, a vanishing $\mathcal{M}_{13}^2$ can only result from an
accidental cancellation between unrelated terms.
We also note that, in contrast to the case of $\mathcal{M}_{12}^2$,
the radiative correction to $\mathcal{M}_{13}^2$ is not enhanced by
$\tan\beta$ with respect to the tree-level part. Thus, its qualitative
impact on our discussion of the alignment conditions is limited, as
long as the scale $Q$ is not too far from $\MS$.

The minimum conditions of the scalar potential can be exploited to
express the mass parameters for the doublets and the singlet in terms
of the other Lagrangian parameters and of the vevs $v_d$, $v_u$ and
$v_S$. In particular, we obtain a relation between the mass parameter
$\tilde{m}_{S R}^{2}$ for the real component of the singlet, see
eq.~(\ref{eq:mSR}), and the matrix element $\mathcal{M}_{13}^2$ given
in eq.~(\ref{eq:alignNMSSMs}): 
\begin{equation}
\tilde{m}_{S R}^{2} ~=~
-\frac{v}{2\,v_S}\, \mathcal{M}_{13}^2
- \frac{v_{S}}{\sqrt 2}
\left(T_{\kappa}+\sqrt 2 \kappa^2 v_{S} + 3 \kappa M_S \right)\,
\label{eq:mSR_vs_M13}~.
\end{equation}

Eq.~(\ref{eq:mSR_vs_M13}) above shows that the condition of vanishing
mixing between $h$ and $S_R$, however realised, carries implications
for the mass of the singlet. The diagonal element for the singlet in the
scalar mass matrix is 
\begin{eqnarray}
\mathcal{M}_{33}^2&=&
\tilde{m}_{S R}^{2} 
\,+\, \frac12 \lambda_S^2 v^2
\,+\, \sqrt2 T_\kappa v_S \,+\, \kappa \left(3 \kappa v_S^2 + 3\sqrt2  M_S v_S
-\lambda_S  s_\beta
c_\beta v^2 \right) \nonumber\\[2mm]
&&+  \frac{3 y_t^2 \lambda_S^2c_\beta^2}{32 \pi^2}  v^2 
\ln\frac{\MS^2}{Q^2}\,+\,(...)~,
\label{eq:M33}
\end{eqnarray}
where we applied the same approximations as in
eq.~(\ref{eq:alignNMSSMs}) for the one-loop correction in the second
line.

We now discuss the simplest case in which the global $U(1)_R$ R-symmetry
is preserved in the superpotential but broken by soft SUSY-breaking
terms, in which case we can set $\kappa$ to zero. Since this implies a
vanishing quartic self-coupling for the singlet, the stability of the
scalar potential requires that we also assume $T_\kappa = 0$, even if the trilinear self-coupling of the singlet
resides in the R-conserving part of the soft SUSY-breaking
Lagrangian. In this scenario, which we shall refer to as the \emph{aligned MDGSSM}, eq.~(\ref{eq:mSR_vs_M13}) shows that the
alignment condition $\mathcal{M}_{13}^2\approx 0$ requires
$\tilde{m}_{S R}^{2}\approx 0$ or $v_S\ll v$. The first of these two
options implies that the CP-even mass eigenstate that is mostly
singlet is relatively light: setting $\tilde{m}_{S R}^{2}= 0$ and
$\kappa=T_\kappa=0$ in eq.~(\ref{eq:M33}), and neglecting the small effect of
the one-loop correction, we find that the value $\lambda_S \approx
0.7$ favored by the alignment condition for the Higgs doublets in the
NMSSM, see eq.~(\ref{eq:alignNMSSMs}), leads to
$\mathcal{M}_{33}^2\approx (122~{\rm GeV})^2$, whereas the value
$\lambda_S \approx 0.25$ implied in our Dirac-gaugino model by the $N
=2$ SUSY relation of eq.~(\ref{eq:SUSYcoupling}) leads to
$\mathcal{M}_{33}^2\approx (44~{\rm GeV})^2$. We remark that the
mixing between $S_R$ and the heavier, non-SM-like scalar $H$, which is
controlled by $\mathcal{M}_{23}^2$, is suppressed when $M_A^2 \gg
\mathcal{M}_{33}^2$, and would in any case lower the mass of the
singlet-like eigenstate.

The definition in eq.~(\ref{eq:mSR}) shows that even the vanishing of
$\tilde{m}_{S R}^2$ requires a cancellation between terms that arise
from different sectors of our $D$-dimensional construction: $m_{1D}$
from the R-conserving boundary superpotential in
eq.~(\ref{GauginomassesSuperpotentials}), $M_S$ from the R-violating
boundary superpotential in eq.~(\ref{R-violatingSuperpotentials}),
$m_S^2$ and $B_S$ from the R-conserving soft SUSY-breaking Lagrangian
in eq.~(\ref{R-conserving-scalar}). If such cancellation is not
realised, the alternative requirement for alignment implied by
eq.~(\ref{eq:mSR_vs_M13}) when $\kappa=0$ is that $v_S \ll v$. This is
not problematic as long as a suitable higgsino mass is provided by the
$\mu$ term in the bulk superpotential, see eq.~(\ref{N=2Lagrangian}).

\bigskip

Finally, the third condition for Higgs alignment in DG models is
${\cal M}^2_{14} = 0$, i.e.~vanishing mixing between $h$ and
$T^0_R$. The formulas for the relevant mass-matrix elements and for
the minimum condition, including the dominant one-loop corrections
from top and stop loops, are similar to
eqs.~(\ref{eq:alignNMSSMs})--(\ref{eq:M33}), with the obvious
singlet-to-triplet replacements but without terms analogous to those
controlled by $\kappa$ in the singlet case:
\begin{eqnarray} 
\label{eq:alignNMSSMt}
  \mathcal{M}_{14}^2 &=& \Delta_{ht}
- \frac{6 y_t \lambda_T c_\beta}{16\pi^2} m_t (A_t - \mu_{\rm eff}\cot\beta)
\ln\frac{\MS^2}{Q^2}\,+\,(...)~,\\
\label{eq:M44}
\mathcal{M}_{44}^2 &=&
\tilde{m}_{T R}^{2} 
\,+\, \frac12 \lambda_T^2 v^2
\,+\,  \frac{3 y_t^2 \lambda_T^2c_\beta^2}{32 \pi^2}  v^2 
\ln\frac{\MS^2}{Q^2}\,+\,(...)~,
\end{eqnarray}
\begin{equation}
\tilde{m}_{T R}^{2} ~=~
-\frac{v}{2\,v_T}\, \mathcal{M}_{14}^2~.
\label{eq:mTR_vs_M14}
\end{equation}
The discussion of the constraints on the triplet mass induced by the
condition of doublet-triplet alignment follows the lines of the
discussion of singlet-triplet alignment for $\kappa=0$, with the
important difference that the condition $v_T \ll v$ must in any case
be satisfied to avoid an excessive contribution to $\Delta\rho$. As a
consequence, it is not necessary to require $\tilde{m}_{TR}^{2}\approx
0$ to obtain approximate alignment. Nevertheless, we remark that the
condition of exact alignment $\mathcal{M}_{14}^2=0$ would imply
$\mathcal{M}_{44}^2 \approx M_W^2$ through
eqs.~(\ref{eq:alignNMSSMt})--(\ref{eq:mTR_vs_M14}).

\subsection{The electroweak fermion sector}

We now outline the mass spectrum of the electroweak fermions, which
will be relevant for our discussion examining the $W$ boson mass.  The
neutralino mass matrix, in the basis
$\tilde{S}, \tilde{B}, \tilde{T}^0, \tilde{W}^0, \tilde{H}_d^0, \tilde{H}_u^0$
reads: 
 \begin{equation}
{\scriptsize \left(\begin{array}{c c c c c c}
M_S + \sqrt{2} \kappa v_S  & m_{DY} & 0     & 0     & - \frac{ \sqrt{2} \lambda_S }{g_Y}M_Z s_W s_\beta &  -  \frac{ \sqrt{2} \lambda_S }{g_Y}M_Z s_W c_\beta  \\
m_{DY} & M_1   & 0     & 0     & -M_Z s_W c_\beta &   M_Z s_W s_\beta  \\
0     & 0     & M_T  & m_{D2} & - \frac{ \sqrt{2} \lambda_T  }{g_2}M_Z c_W s_\beta & - \frac{ \sqrt{2} \lambda_T  }{g_2}M_Z c_W c_\beta  \\
0     & 0     & m_{D2} & M_2   &  M_Z c_W c_\beta & - M_Z c_W s_\beta  \\
-\frac{ \sqrt{2} \lambda_S }{g_Y}M_Z s_W s_\beta & -M_Z s_W c_\beta & -\frac{ \sqrt{2} \lambda_T  }{g_2}M_Z c_W s_\beta &  M_Z c_W c_\beta & 0    & -\mueff \\
-\frac{ \sqrt{2} \lambda_S }{g_Y}M_Z s_W c_\beta &  M_Z s_W s_\beta & -\frac{ \sqrt{2} \lambda_T  }{g_2}M_Z c_W c_\beta & -M_Z c_W s_\beta & -\mueff & 0    \\
\end{array}\right) }
\label{diracgauginos_NeutralinoMassarray}
\end{equation}

The chargino masses, $- \frac{1}{2} ( (v^-)^T \mathcal{M}_{\chi^\pm} v^+ + h.c.)$
in the basis $v^+ = (\tilde{T}^+,\tilde{W}^+,\tilde{H}^+_u)$, 
$v^- = (\tilde{T}^-,\tilde{W}^-,\tilde{H}^-_d)$,
are given by
\begin{equation}
\mathcal{M}_{\chi^\pm} = 
\left(\begin{array}{c c c}
M_T   & m_{D2} + g_2 v_T & \lambda_T v  c_\beta \\
m_{D2} - g_2v_T & M_2   & \frac{g_2 v}{\sqrt{2}} s_\beta \\
-\lambda_T v s_\beta & \frac{g_2 v}{\sqrt{2}} c_\beta & \mueff - \sqrt{2}\lambda_T v_T \\
\end{array}\right) 
\label{diracgauginos_CharginoMassarray}
\end{equation}
%
%
%
and do not depend on $\kappa$, but we have written for completeness
the Majorana gaugino masses $M_{1,2}$ for the bino and wino
respectively.

\subsection{Scenarios}
\label{sec:scenarios}

We have described the general features of minimal Dirac Gaugino models, and explained how some values of the couplings allow Higgs alignment to automatically occur at tree-level. Below, we shall consider the following different scenarios corresponding to specific choices of the model parameters:

\begin{itemize}
\item{\bf General MDGSSM}

  In the general MDGSSM, the only source of $R$-symmetry violation comes from a small $B_\mu$ term, which is a radiatively stable condition (the renormalisation group running will not generate other $R$-symmetry violating terms from a $B_\mu$ term). This excludes all of $W_{NR}$; the only supersymmetric parameters beyond those of the MSSM retained are $\lambda_S, \lambda_T$, but those are allowed to have any value. Supersoft  masses \cite{Fox:2002bu} are allowed, and soft supersymmetry-breaking masses are allowed, but squark/sfermion trilinears and $T_S, T_T$ are not. On the other hand, trilinears involving the adjoint scalars may be allowed (so $T_\kappa, T_{ST}, T_{SO}$) but may be argued to be small in typical models. The couplings $\lambda_S, \lambda_T$ enhance the Higgs mass and W mass at the same time; while large values of $\lambda_T$ were previously deemed problematic for the $\rho$ parameter, they are now a virtue.


\item{\bf MSSM without $\mu$ term (RIP)}
  
This model, described in \cite{Nelson:2002ca}, is identical to the general MDGSSM execpt that the $\mu$-term is set to zero (although a small $v_S$ generates a tiny effective $\mu$). In the MSSM this would yield massless higgsinos, but here the higgsinos obtain a mass through $\lambda_T$ which causes them to mix with the triplet fermion. Thus the $\mu$ problem of the MSSM is solved, at the expense of an upper bound on the chargino masses, which, as we shall see, leads to the model being ruled out.

\item{\bf Aligned MDGSSM}

By taking  $\lambda_S, \lambda_T$ to their their $N=2$ values given in Eq. (\ref{eq:SUSYcoupling}) in the general MDGSSM we guarantee alignment with the heavy Higgs at tree level. If we further impose that $\Delta_{hs} = 0$, as described above,  then the singlet also has negligible mixing with the SM-like Higgs. We shall refer to this scenario as the \emph{aligned MDGSSM}. 

\item{\bf DGNMSSM}

If we instead allow R-symmetry violation in the superpotential and the associated soft-breaking trilinears -- in particular for $\kappa, T_\kappa$ and $T_S$ -- we can generate $\mu$ and $B_\mu$ terms through a substantial expectation value for the singlet. The model thus resembles the NMSSM, especially if we set the $\mu, B_\mu$ terms (and $M_S, M_T$) to zero; this was proposed in \cite{Benakli:2011kz}. We shall therefore refer to this scenario as the DGNMSSM. 

\item{\bf Aligned DGNMSSM}

We can achieve aligment in the DGNMSSM by setting $\lambda_S, \lambda_T$ to their $N=2$ values and taking $\Delta_{hs} =0$. We shall refer to this scenario as the aligned DGNMSSM; in this work, when we consider the W boson mass, we shall also enforce $\Delta_{Hs} =0$ which guarantees that the singlet couplings to SM fields are small, rendering light singlets safe from collider searches.

\end{itemize}

\section{W mass in Dirac Gaugino models}
\label{SEC:MW}


Dirac gaugino models offer two methods of explaining an enhancement of the W boson mass with respect to the SM: either quantum corrections or a tree-level expectation value for the triplet scalar. In the case of quantum corrections, there are new contributions to the W mass compared to the MSSM, again coming from interactions related to the adjoint triplet; in particular the coupling $\lambda_T$. 

Recall the definition of the $\rho$ parameter is $M_W^2 \equiv \rho c_W^2 M_Z^2;$ through the presence of the adjoint triplet, DG models contain a tree-level modification to this relation compared to the SM:
\begin{align}
\rho \equiv 1 + \Delta \rho_{\rm tree} + \Delta \rho = 1 + 4 \frac{v_T^2}{v^2} + \Delta \rho.
\end{align}
One could consider a modification to $c_W$ instead of $\rho$ as an explanation of an enhanced $W$ mass, but this is discounted based on electroweak precision tests (see e.g. \cite{Du:2022fqv}); on the other hand, a triplet expectation value is one of the most generic and acceptable ways of enhancing the W mass at tree level. Naively we could then take the observed value of $M_Z$ and the standard value of $c_W$ and infer the value of $\Delta \rho$ to obtain a given value of $M_W$. However, in the SM and in any BSM theory it is necessary to take certain electroweak observables as input, and in our setup we will take the conventional choice of the Z mass, $G_F$ and $\alpha$. When we modifiy $\rho$ this gives both a modification to $M_W$ \emph{and} a small modification to $\sin^2 \theta_W$. At tree-level this is
\begin{align}
  \Delta_{\rm tree} M_W^2 =& \frac{c_W^2}{c_W^2 - s_W^2} (M_W^2)_{\rm SM} \Delta \rho_{\rm tree}, \nn\\
  \Delta_{\rm tree} s_W^2 =& - \frac{s_W^2 c_W^2}{c_W^2 - s_W^2} \Delta \rho_{\rm tree},
\end{align}
so if we want to obtain $M_W = 80.424$ MeV we need $\Delta M_W^2 = 11\, (\mathrm{GeV})^2.$ Taking $s_W^2 =0.23121$  this gives
\begin{align}
  \Delta^{\rm CDF} \rho_{\rm tree} =& 0.0012 \nn\\
  \Delta_{\rm tree}^{\rm CDF} s_W^2 =& - 4 \times 10^{-4}.
\end{align}
Interpreted as a tree-level expectation value for the triplet, this yields
\begin{align}
v_T \simeq& 4\ \mathrm{GeV},
\end{align}
which was previously at the upper bound of what was acceptable. 

In the minimal Dirac gaugino model, we have from equation (\ref{eq:mTR_vs_M14})
\begin{align}
\tilde{m}_{TR}^2 = - \frac{v}{2v_T} \Delta_{ht} = - \frac{\Delta_{ht}}{\sqrt{\Delta \rho_{\rm tree}}}.   
\end{align}
For a small $v_T$, if we do not tune $\Delta_{ht} \approx 0$, then triplets must be heavy. However, we also need heavy winos to evade collider bounds and this implies large $m_{D2}$: the connection between electroweakino masses and the expectation value of the triplet is a novel feature of this class of models.  In \cite{Goodsell:2020lpx} the conclusion was that for winos above $700 \GeV$ there were essentially no constraints on the higgsinos beyond LEP. If the contribution from $m_{D2}$ dominates $\Delta_{ht}$ this implies
\begin{align}
m_{T}^2 \sim& - \frac{g_2 v m_{D2} c_{2\beta}}{\sqrt{\Delta \rho_{\rm tree}}} \sim (1.8\ \mathrm{TeV})^2 \times \left(\frac{700\ \GeV}{m_{D2}} \right) \times \left( \frac{c_{2\beta}}{-1}\right) \times \sqrt{\frac{0.0012}{\Delta \rho}},   
\end{align}
which is a natural scale for supersymmetric scalars. Of course, we can have lighter winos provided that the neutralino is not too light, above around $200$ to $300$ GeV \cite{Goodsell:2020lpx}.

However, the model also contains ample room for quantum corrections to also enhance the W mass. In the following, we shall investigate this for the different scenarios described in section \ref{sec:scenarios}.


\subsection{Numerical setup}

In order to accurately compute the quantum corrections to the W mass, we use a new EFT approach, closely related to that of \cite{Athron:2022isz}, implemented in the spectrum-generator-generator \SARAH. We use the expression:
\begin{align}
  M_W^2 =& (M_W^2)_{SM} \bigg( 1 + \frac{s_W^2}{c_W^2 - s_W^2} \bigg[ \frac{c_W^2}{s_W^2} (\Delta \rho_{\rm tree} + \Delta \rho) - \Delta r_W - \Delta \alpha\bigg]\bigg)
\end{align}
where $(M_W^2)_{SM}$ is the full two-loop W mass in the SM, as computed in \cite{Degrassi:2014sxa}, depending on the pole masses of the Higgs boson, top quark, $\alpha_s$ and $\Delta \alpha_{\rm had}^{(5)}$. We use the interpolating function from that paper. When we use the average values of the Higgs mass of $125.09$ GeV and the top quark mass of $172.89$ GeV this function gives us $M_W = 80.354$ GeV, just 2 MeV higher than the current world average for the W boson mass in the SM.
The expressions computed in the square brackets are the \emph{differences} between the high-energy theory (HET) and the SM:
\begin{align}
  \Delta \rho \equiv& \mathrm{Re} \bigg[ \frac{\Pi_{ZZ}^{HET}(M_Z^2)}{M_Z^2} - \frac{\Pi_{WW}^{HET}(M_W^2)}{M_W^2} \bigg] - \mathrm{Re} \bigg[ \frac{\Pi_{ZZ}^{SM}(M_Z^2)}{M_Z^2} - \frac{\Pi_{WW}^{SM}(M_W^2)}{M_W^2} \bigg] \nn\\
  \Delta r_W \equiv& \bigg[\frac{\Pi_{WW}^{HET}(0)}{M_W^2} -  \frac{\Pi_{WW}^{HET}(M_W^2)}{M_W^2} + \delta_{VB}^{HET} \bigg] - \bigg[\frac{\Pi_{WW}^{SM}(0)}{M_W^2} -  \frac{\Pi_{WW}^{SM}(M_W^2)}{M_W^2} + \delta_{VB}^{SM}\bigg].
\end{align}
$\Delta \alpha$ are now the gauge threshold corrections between the HET and the SM for the electromagnetic gauge coupling divided by $\alpha$ (so they do not now depend on $\Delta \alpha_{\rm had}^{(5)}$).

In {\tt SARAH}, we compute the expression in square brackets at the matching scale, which is the mass of the heavy particles. We are therefore ignoring the running from that scale down to the electroweak scale, which is of controllable size, but will nevertheless be included in a future development of the code. One could argue that we should instead perform the matching at the electroweak scale, but then all of the loop functions will contain large logarithms and there can be larger, spurious, running of the couplings of the high-energy theory which can spoil the results.

In order to be a strict one-loop matching between the HET and the SM, the weak mixing angle in the above must be the $\ov{MS}$ or $\ov{DR}$ value at the matching scale $Q$:
\begin{align}
s_W^2 =& \frac{g_Y^2 (Q)}{g_2^2(Q) + g_Y^2(Q)}.
\end{align}
To extract this, we match the calculations of the Z-boson mass, $\alpha(Q)$ -- and we also compute the decay of the muon at the matching scale. In practice, this means that we extract the couplings in the SM at the top mass scale without including the effects of new physics, then run them up to the matching scale. The threshold corrections to $\alpha$ between the two theories are simple to compute since it is unbroken and yield $\Delta \alpha$; the coupings $g_Y(Q), g_2(Q)$ in the high-energy theory are chosen to solve the equation
\begin{align}
c_W^2 s_W^2 =& (c_W^2 s_W^2)_{\rm SM} \times \frac{(1 + \Delta \alpha + \Delta \hat{r})}{1 + \Delta \rho_{\rm tree}}
\end{align}
where
\begin{align}
\Delta \hat{r} \equiv& \bigg[(1 +  \Delta \rho_{\rm tree})\frac{\Pi_{WW}^{HET}(0)}{M_W^2} -  \frac{\Pi_{ZZ}^{HET}(M_Z^2)}{M_Z^2} + \delta_{VB}^{HET} \bigg] - \bigg[\frac{\Pi_{WW}^{SM}(0)}{M_W^2} -  \frac{\Pi_{ZZ}^{SM}(M_Z^2)}{M_Z^2} + \delta_{VB}^{SM}\bigg],
\end{align}
which is done iteratively by progressively running up and down and updating at each step, along with all the other quantities in the high-energy theory. The value for $ (c_W^2 s_W^2)_{\rm SM}$ includes a compensatory term for corrections from $\ov{\mathrm{MS}}$ to $\ov{\mathrm{DR}}$ if needed.

In the \SARAH model file, the couplings $\lambda_S,\lambda_T, T_S, T_T $ are defined differently to the above: we have ${\tt lam} = - \lambda_S, {\tt LT} = \sqrt{2} \lambda_T, {\tt T[lam]} = - T_S, {\tt T[LT]} = \sqrt{2} T_T$. Hence in our plots and benchmark points we list the values in terms of $- \lambda_S, \sqrt{2} \lambda_T, - T_S$, which makes the correspondence with the numerical codes exact.

\subsection{MSSM without $\mu$ term}
\label{sec:MSSMnomu}

The ``MSSM without $\mu$ term'' proposed in \cite{Nelson:2002ca} was an intriguing solution to the $\mu$ problem. 
It was however challenged by the requirement of having a high enough Higgs mass, chargino mass and not too large $\rho$; indeed the lack of intersection of points satisfying the latter two was demonstrated in  \cite{Benakli:2012cy}. It might therefore be tempting to revisit this model in light of the new data about the $W$ mass. However, we shall demonstrate here that it is conclusively ruled out.
                                         
Putting aside the Higgs mass constraint, the see-saw effect on the charginos is a problem. LEP put a lower limit on the mass of the lightest chargino of 94~GeV \cite{ALEPH:2002gap,Abdallah:2002aik,LEPchargino}. In that model the chargino mass is
\begin{equation}
\mathcal{M}_{\chi^\pm} \underset{v_T \simeq 0}{\longrightarrow} 
\left(\begin{array}{c c c}
0 & m_{D2} &\frac{ {2} \lambda_T  }{g_2} M_Z c_W c_\beta \\
m_{D2}  & 0  & \sqrt{2} M_Z c_W s_\beta \\
- \frac{ {2} \lambda_T  }{g} M_Z c_W s_\beta & \sqrt{2} M_Z c_W c_\beta & 0 \\
\end{array}\right) 
\end{equation}
It is known that it is possible to fulfil the LEP bound by a careful choice of $\lambda_T$ and $m_{D2}$: a large value of $\lambda_T$ as well as $m_{D2}$ around 107~GeV is needed to maximize the mass of the lightest chargino. It would then be made of a higgsino-wino mixture with \emph{two} charginos that are light and one (wino-like) somewhat heavier. Unfortunately, subsequent LHC searches are especially sensitive to winos up to about $800$ GeV, see \cite{Goodsell:2020lpx}. It might be possible to evade this constraint if the light wino and neutralino are close enough in mass so that decays such as $\tilde{\chi}^\pm \rightarrow \tilde{\chi}^0 + W^\pm$ are not possible. Most likely this is difficult or impossible to achieve, but without a detailed investigation we cannot exclude the possibility that some region of parameter space might evade direct LHC searches; we can only apply the LEP constraint as a hard lower bound on the chargino mass. 

\begin{figure}\centering
  \includegraphics[width=0.5\textwidth]{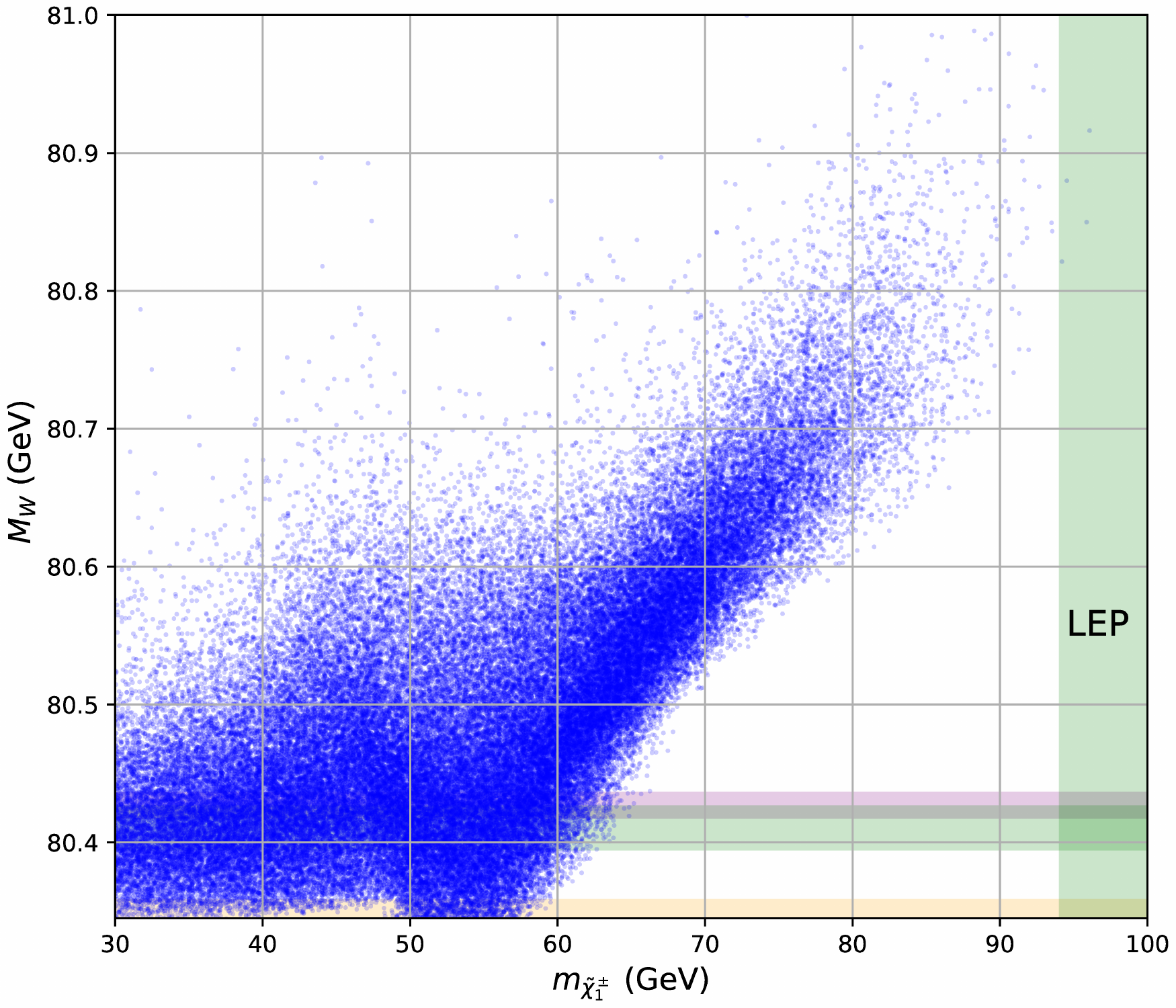}
 \caption{\label{FIG:nomu} W boson mass vs lightest chargino mass for points generated in the `MSSM without $\mu$-term.' The horizontal bands show the SM, Tevatron and world average masses for the W boson mass; the vertical green band shows the LEP constraint on the lightest chargino.}
\end{figure}

In \cite{Benakli:2012cy} it was demonstrated that that the corrections to $\Delta \rho$ were correlated with the lightest chargino mass; in order to evade the LEP bound, $\lambda_T$ has to be large and this drives large $\Delta \rho$. Here we can give a striking confirmation of this observation by plotting the $W$ mass against the lightest chargino mass  for a sample of $\mathcal{O}(70000)$ spectra generated using \SARAH. We fix the octet scalar, and all squark and slepton soft masses via $m_O^2 = m_{\tilde{q}}^2 = m_{\tilde{l}}^2 = 10 $ TeV${}^2$, and fix the gluino mass to $3$ TeV; this ensures that they are beyond all current and near-future bounds. Then we scan over the ranges:
\begin{align}
  m_{DY} \in [100,500]\, \GeV,& \quad m_{D2} \in [100,250]\, \GeV, \quad v_S \in [-50, 50]\, \GeV, \quad v_T \in [-5, 5]\, \GeV \nn\\
  B_\mu \in [10^4, 10^6]\, (\GeV)^2,& \quad \lambda_T \in [-1,1], \quad \lambda_S \in [-1.5,1.5], \quad \tan \beta \in [2,50].
\end{align}
We use a Markov Chain Monte-Carlo (MCMC) algorithm to generate points (this helps to obtain points with non-tachyonic spectra with Higgs mass close to the observed value compared to a random scan). There is not intended to be a genuine statistical interpretation of the distribution of the points, but the overall envelope should show where valid sets of parameters exist. The results are shown in figure \ref{FIG:nomu}, where the LEP constraint is shown as a vertical green band. It can be clearly seen that it is not possible to both satisfy the LEP constraint and have an acceptable value for the W boson mass; from the W boson mass alone we would predict a chargino of mass below $65$ GeV.

\subsection{W mass in the MDGSSM}
\label{sec:WMDGSSM}

In the minimal Dirac Gaugino extension of the Standard Model (MDGSSM) the most typical scenario is to assume that R-symmetry is broken only via a $B_\mu$ term, i.e. the Higgs sector is special. It is identical to the previous model except that we allow a $\mu$-term. This, however, makes all the difference: now the higgsino mass is not bounded from below, not requiring a large mixing with the winos; and further the enhancement to the Higgs mass is under control. 

We perform a new MCMC scan with parameters allowed to vary within the ranges:
\begin{align}
  m_{DY} \in [100,700]\, \GeV, &\quad m_{D2} \in [100,1200]\, \GeV, \quad v_S \in [-50, 50]\, \GeV, \quad v_T \in [-5, 5]\, \GeV \nn\\
  \mu \in [0, 1000]\, \GeV,& \quad B_\mu \in [10^4, 10^6]\, (\GeV)^2, \nn\\
  \quad \sqrt{2}\lambda_T \in [-1.5,1.5],& \quad \lambda_S \in [-1.5,1.5], \quad \tan \beta \in [2,50].
\end{align}
We fix the octet scalar, and all squark and slepton soft masses via $m_O^2 = m_{\tilde{q}}^2 = m_{\tilde{l}}^2 = 10 $ TeV${}^2$, and fix the gluino mass to $3$ TeV. We choose a likelihood function to be a product of a gaussian in the Higgs mass with mean $125$ GeV and standard deviation $3$ GeV, a gaussian in the W mass with mean $80.413$ GeV and standard deviation $20$ MeV, and a sigmoid on the constraints (given as the maximum ratio of predicted cross-section to observed, across all channels) from \HiggsBounds \cite{Bechtle:2008jh,Bechtle:2011sb,Bechtle:2013wla,Bechtle:2020pkv}, which strongly suppresses the likelihood when the observed cross-section ratio is greater than one, but is otherwise close to unity. This choice of likelihood function is merely a device to select desirable points, and the distribution of the points is not meant to have a statistical interpretation in terms of their Bayesian likelihood; in particular, the theory uncertainty on the Higgs mass does not have a statistical interpretation, and a window of $3$ GeV is a conservative estimate of the average error, since we use the latest two-loop corrections in the generalised effective potential and gaugeless limit \cite{Goodsell:2015ira,Braathen:2016mmb,Braathen:2016cqe,Braathen:2017izn} with pole-mass matching onto the SM \cite{Staub:2017jnp} (see \cite{Slavich:2020zjv} for a recent review). Such a conservative window of $3$ GeV is employed because only a relatively small proportion of points actually generate a spectrum, and in principle a two-stage procedure pre-filtering points along those suggested in \cite{Goodsell:2020lpx,Goodsell:2022beo} would probably be more efficient -- or in addition the ability to invert the vacuum minimisation relations and compute $v_S, v_T$ instead of treating them as inputs, but this is not yet automatically possible in the code in a way that would correctly incorporate the loop corrections to the Higgs masses \cite{Braathen:2021fyq}.

\begin{figure}\centering
  \includegraphics[width=0.45\textwidth]{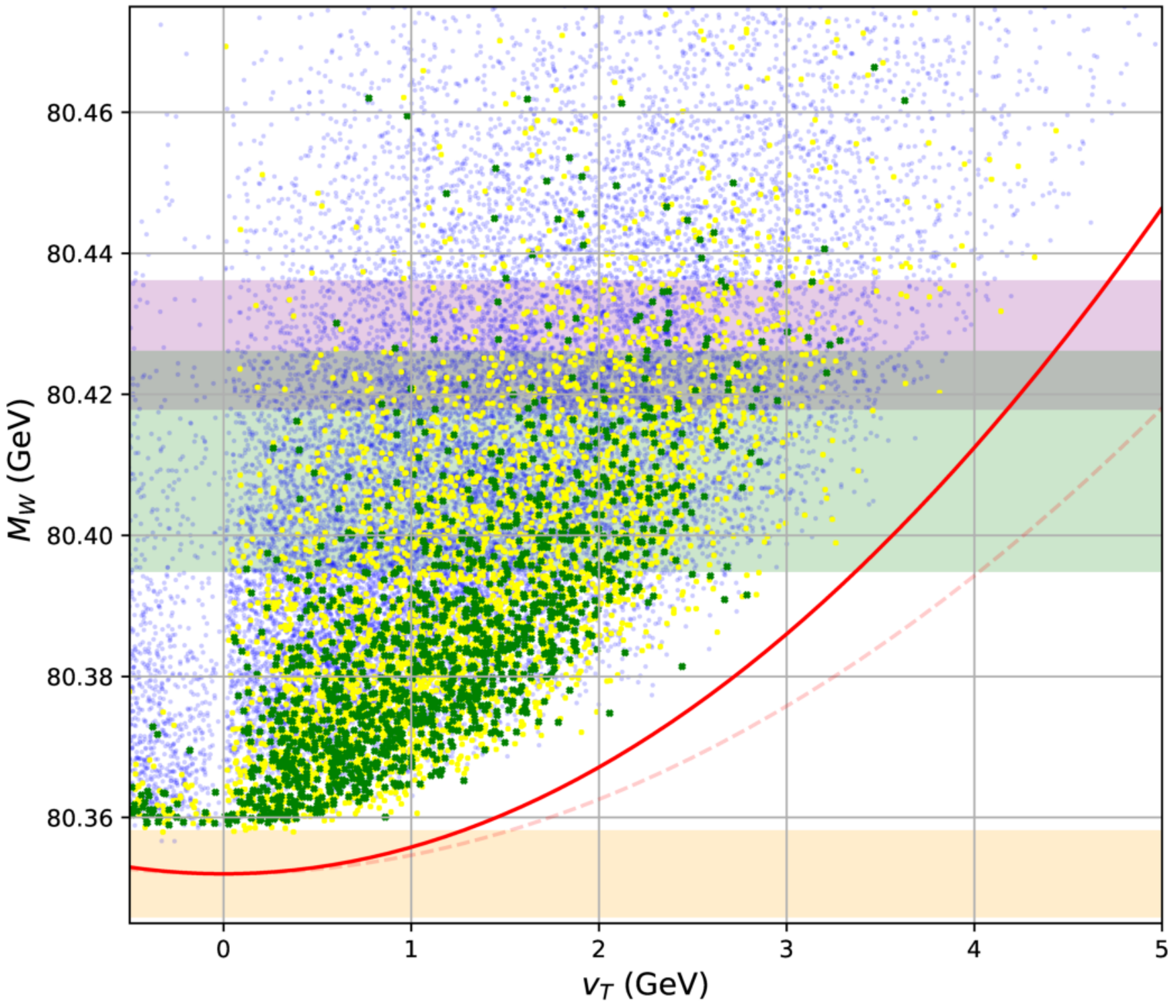} \includegraphics[width=0.45\textwidth]{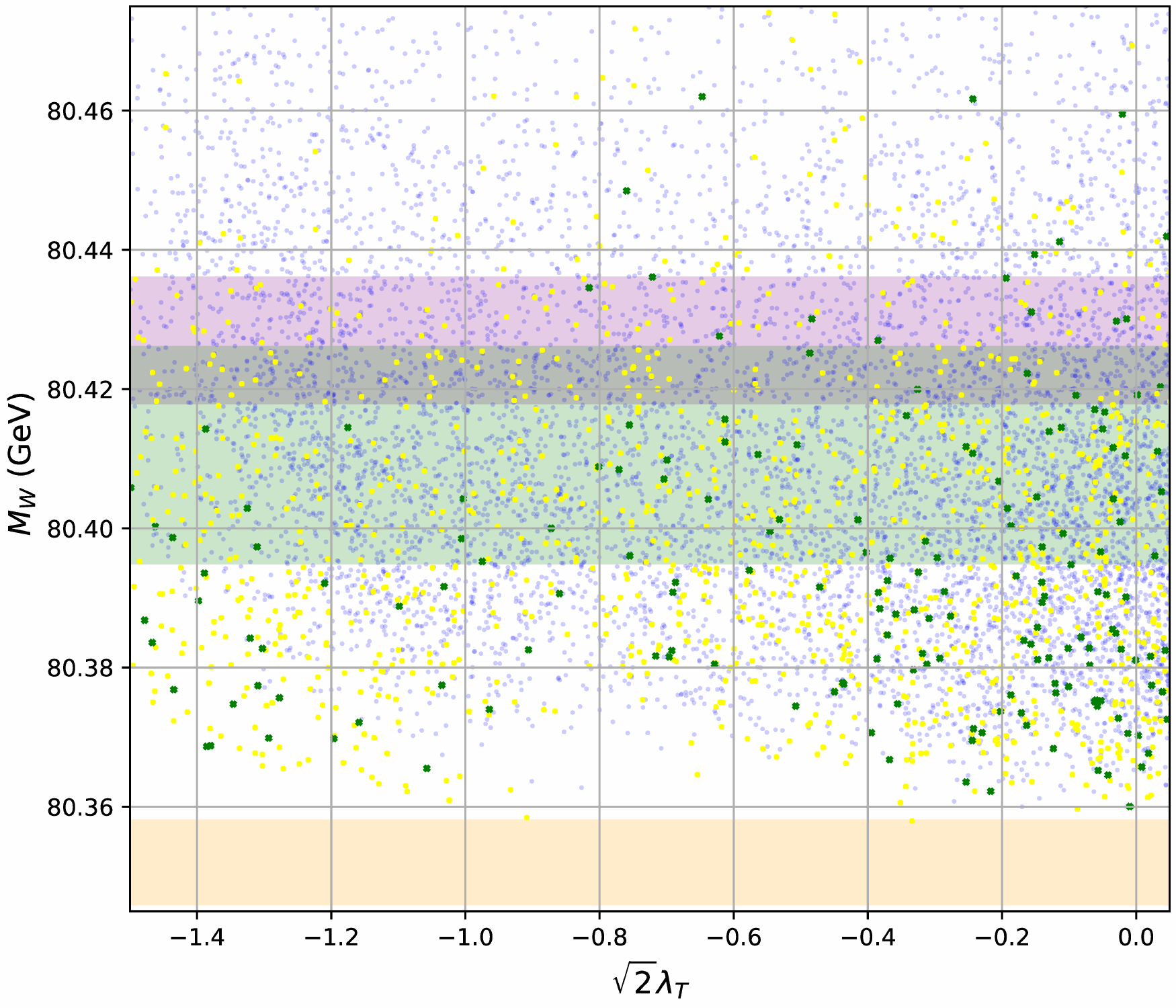}
 \caption{\label{FIG:MDGSSM} Left:  W boson mass vs $v_T$ in the MDGSSM, with the red curve showing the tree-level prediction. Right:  W boson mass vs $\sqrt{2} \lambda_T$ in the MDGSSM. Colours of the points are described in the text, with those obeying the strictest cuts shown in green. The colourful horizontal bands show the SM range in light orange; the Tevatron+LEP average in purple, and a conservative world average in green.}
\end{figure}

\begin{table}\centering
  \begin{tabular}{|c|c|c|c|}\hline \hline
  & MDG1 & MDG2 & MDG3 \\ \hline \hline
$m_{DY}$\, (GeV) & 280 & 285 & 245\\
$m_{D2}$\, (GeV) & 983 & 941 & 940\\
$\mu$\, (GeV) & 276 & 255 & 353\\
$\tan \beta$ & 48 & 46 & 47\\
$-\lambda_S$ & 1.179 & 1.074 & 1.112\\
$\sqrt{2}\lambda_T$ & -0.487 & 0.502 & 0.099\\
$B_\mu$\, $(\mathrm{GeV})^2$ & 828838 & 794477 & 938787\\
$v_S$\, (GeV) & 5.0 & 4.3 & 1.4\\
$v_T$\, (GeV) & 2.3 & 2.7 & 2.8\\\hline\hline
$m_{h_1}$\, (GeV) & 125.4 & 124.7 & 124.7\\
$m_{h_2}$\, (GeV) & 3120.9 & 2274.3 & 2405.8\\
$m_{A_1}$\, (GeV) & 2394.2 & 1221.4 & 1456.5\\
$m_{H_1^\pm}$\, (GeV) & 2400.2 & 1213.0 & 1455.3\\
$m_{\tilde{\chi}_1^0}$\, (GeV) & 217.9 & 231.5 & 233.0\\
$m_{\tilde{\chi}_1^\pm}$\, (GeV) & 255.0 & 275.9 & 362.0\\
$m_{W}$\, (GeV) & 80.425 & 80.420 & 80.421\\\hline\hline
\end{tabular}
\caption{\label{BenchmarksMDG} Benchmark points for the MDGSSM. Input parameters are given above the double line, and masses of the most important particles below. }
\end{table}

We show plots in figure \ref{FIG:MDGSSM} for $M_W$ against $v_T$ (left plot), and then for $M_W$ against $\lambda_T$ (right plot), to show the points that benefit from large quantum corrections as the means of enhancing the $W$ mass: the tree-level expectation just from modifying $v_T$ is shown as a solid red curve on the left plot.  We also show as a dashed red curve the value of $M_W$ that would be obtained by insisting that the shift in $\Delta \rho$ only modifies $M_W$ without changing $\sin \theta_W$ (if a different method of matching onto the SM parameters were used, for example). 

All points shown satisfy all Higgs bounds; have a charged Higgs heavier than $600$ GeV (so are safe from $B \rightarrow s \gamma$ constraints \cite{Misiak:2017bgg}); have charginos heavier than the LEP limit and winos heavier than $600$ GeV. These baseline selections are marked as blue points in the plots; there are about 36000, of which about 10000 have $|v_T| < 1$ GeV. Points shown in yellow further have charginos heavier than $200$ GeV, $m_{D2} > 700$ GeV; while those marked in green have charginos heavier than $250$ GeV and $m_{D2} > 800$ GeV (about $1600$ points in our sample survive these cuts). The green points are thus almost certainly guaranteed to be safe from current collider bounds (although they may yet be probed in future). The requirement of heavy charged Higgs scalars sets the MSSM-like neutral and pseudoscalar masses to be heavy, and essentially guarantees the safety of all selected points from constraints on the couplings of the SM-like Higgs. Nevertheless, we also filtered the green points with constraints from {\tt HiggsSignals} \cite{Bechtle:2013xfa,Bechtle:2020uwn}. The different categories of points show the expected wider range of enhancements to the W mass as the charginos become lighter. We provide a selection of benchmark points, with the input parameters and crucial data, in table \ref{BenchmarksMDG}.

The first clear observation is that the quantum corrections to the W mass are at least as important as the tree-level contribution from the expectation value $v_T$, and the generic contribution to the W mass is positive with no points below the red curve. The asymmetry of the plot with $v_T$ is due to the fact that we only take positive Dirac gaugino masses. It is important to note that in the red curves we take the SM value of the W boson mass to be $80.352$ GeV, whereas the fitting function of \cite{Degrassi:2014sxa} as employed in \SARAH gives a value of $80.354$ GeV for the central values; and gives $80.356$ for a Higgs mass at the lower bound of our permitted range of $122$ GeV. However, it is clear that the quantum corrections in our sample are generally more important than $\Delta \rho_{\rm tree}$. Indeed, with the parameter ranges we have chosen, we are not within the range of masses required for decoupling of the quantum corrections (we have checked that the quantum corrections to $M_W$ smoothly drop to near zero as the masses of all particles are raised to about $2$ TeV or higher). 

The second observation is that there is no clear correlation between the $W$ mass and the parameter $\lambda_T$. Due to our requirement of a large wino mass, the selected points have light neutralinos/charginos of mixed bino/higgsino type. The mass splitings among the higgsinos -- and thus the contribution to the $W$ mass -- can be driven large by $\lambda_S$ and $\lambda_T$. These couplings also enhance the Higgs mass at tree-level, so large values are favoured in the scans because we fixed the stop masses at $\sqrt{10}$ TeV. In this class of model there is therefore no particular preference for one or the other coupling. 

Since the selected points generally have a mixed bino/higgsino LSP, they have good dark matter candidates, but it is expected that the relic density should be underdense. A detailed investigation of the dark matter-collider complementarity for scenarios satisfying the latest W mass data along the lines of \cite{Goodsell:2020lpx} would be an interesting subject for future work, provided that the latest LHC analyses can be recast. As mentioned above, the possibility of a large $\lambda_T$ (and the presence of the singlet scalar/fermion) distinguishes higgsinos in this scenario from those in the MSSM. It is clear that this class of models provides a very natural explanation for an enhancement to the W mass compared to the Standard Model.

\subsection{W mass in the aligned MDGSSM}

\begin{figure}\centering
  \includegraphics[width=0.45\textwidth]{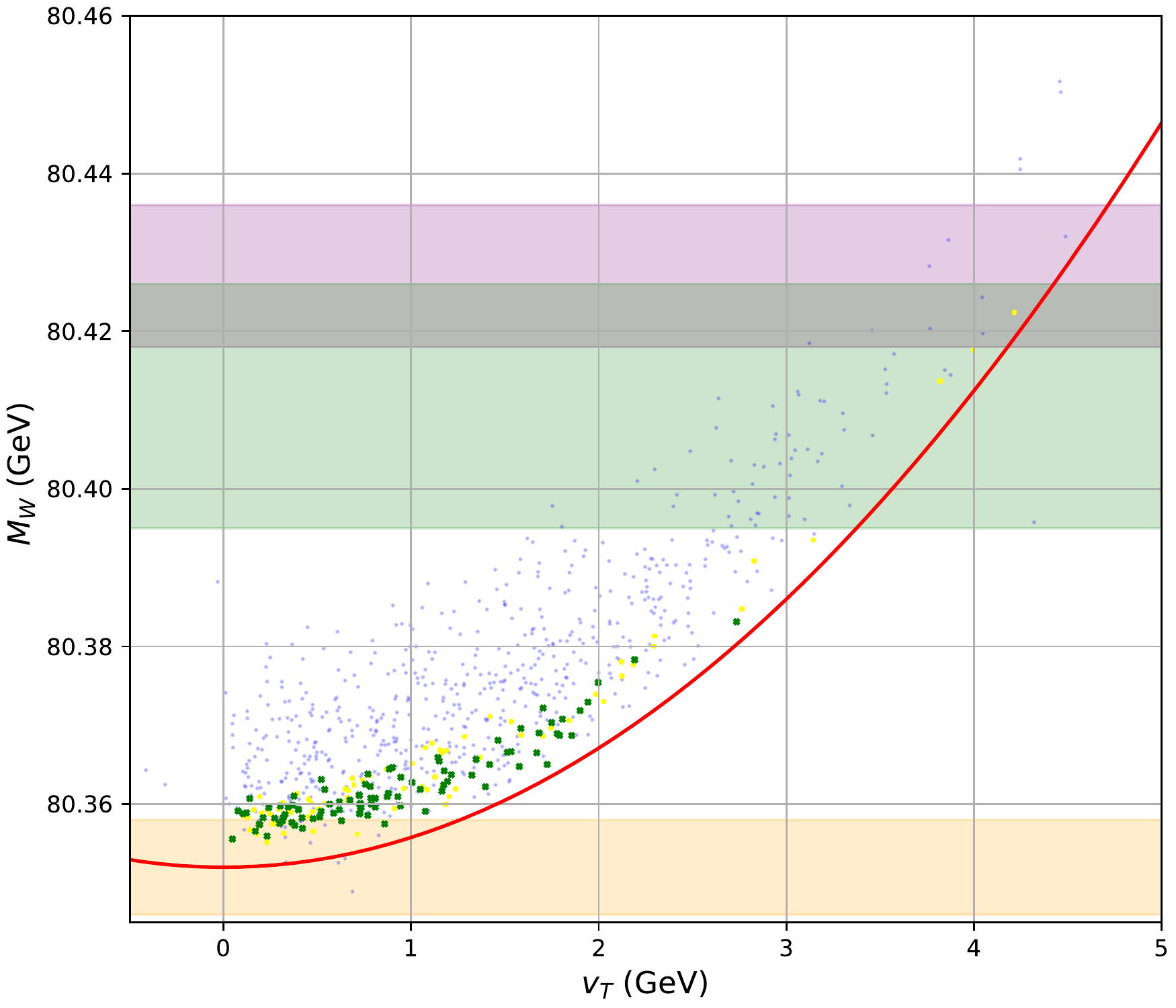} \includegraphics[width=0.45\textwidth]{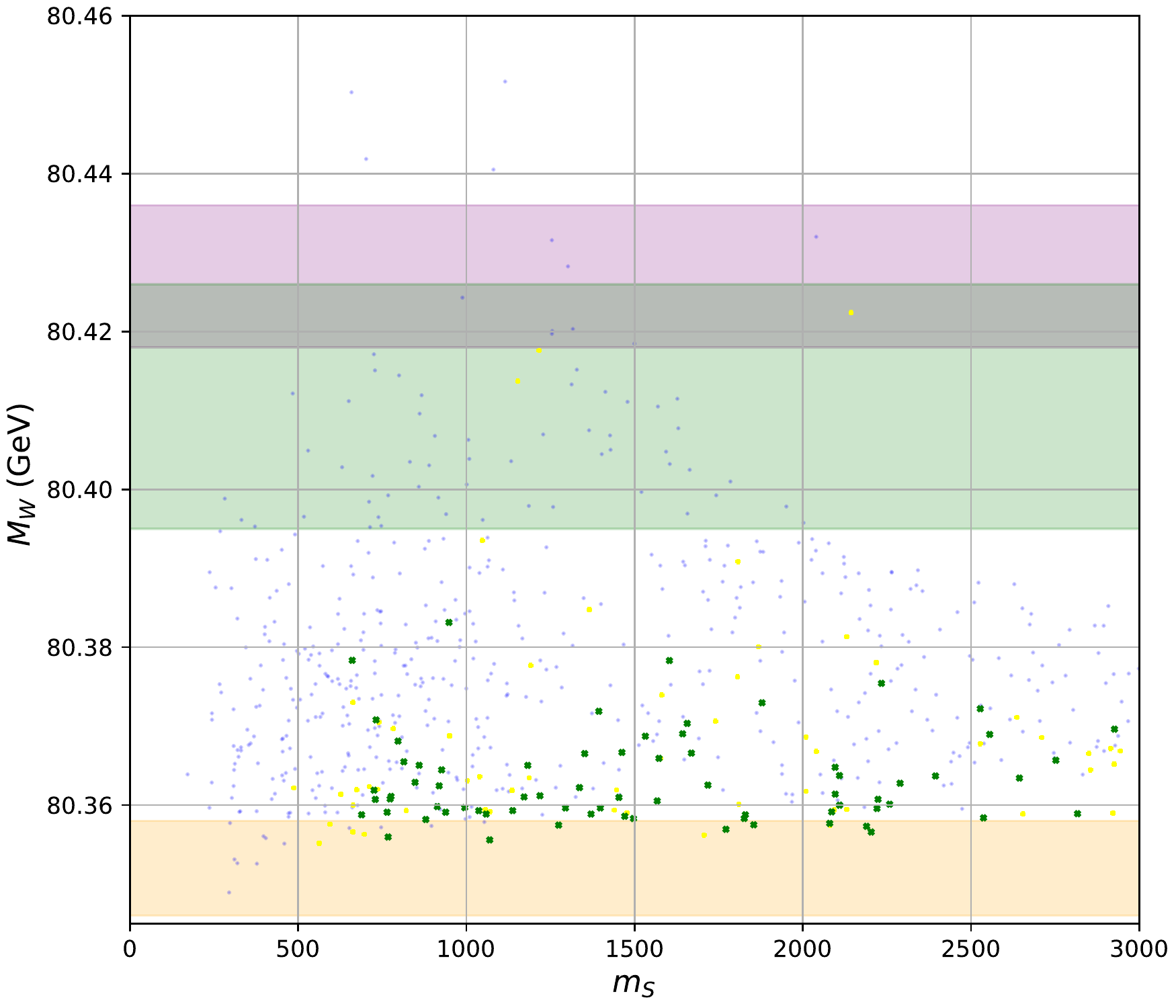}
 \caption{\label{FIG:AMDGSSM} Points in the aligned MDGSSM. Left: $M_W$ against  triplet expectation value. Right: $M_W$ against singlet-like scalar mass. Colour coding as for previous plots.}
\end{figure}

In the aligned MDGSSM (where the only source of R-symmetry breaking is the $B_\mu$ term, and we take $T_T =0$), we choose the parameters to induce alignment at tree-level in the MDGSSM (so $\lambda_S=g_Y/\sqrt{2}, \lambda_T = g_2/\sqrt{2}$ and $m_{DY} = c_{2\beta} \mueff$ ) but taking $T_S = 0$ (which means the mixing of the singlet with the heavy Higgs cannot vanish unless it is heavy). To study this scenario we perform a scan with the same strategy as before except that now, since $\lambda_S, \lambda_T$ are fixed, it is necessary to vary the masses of the stops/sbottoms to allow us to find the observed value of the Higgs mass; there is also therefore a preference for models with larger $\tan \beta$ since the tree-level contributions to the Higgs mass at low $\tan \beta$ are not sufficient. We therefore use a common mass for the third generation squarks $M_{\rm SUSY};$ we set $m_{Q, 33}^2 = m_{U,33}^2 = m_{D,33}^2 = M_{\rm SUSY}^2$ (the other squarks and sleptons we retain fixed at $\sqrt{10}$ TeV). The remaining parameters are scanned via the same MCMC algorithm in the ranges:
\begin{align}
  m_{D2} \in [400,1500]\, \GeV,& \quad v_S \in [-250, 250]\, \GeV, \quad v_T \in [-5, 5]\, \GeV \nn\\
  \mu \in [-1000, 1000]\, \GeV,& \quad B_\mu \in [10^4, 10^6]\, (\GeV)^2, \nn\\
 M_{\rm SUSY}\in [2,10]\, \mathrm{TeV},& \quad \tan \beta \in [2,50].
\end{align}
We mostly find points with small $\mu/m_{DY}$ and very little enhancement to the W mass because the electroweakinos tend to be light. We give benchmark points in table \ref{TAB:AMDGSSM} and plots in figure \ref{FIG:AMDGSSM} which demonstrate the lack of enhancement and scarcity of points (790 survived from a scan for one million).

\begin{table}\centering
\begin{tabular}{|c|c|c|c|c|c|c|} \hline \hline
  & AMDG1 & AMDG2 & AMDG3 & AMDG4 & AMDG5 & AMDG6 \\ \hline \hline
$M_{\rm SUSY}$\, (GeV) & 8636.2 & 4550.9 & 5181.9 & 8436.1 & 7357.2 & 5454.9\\
$m_{D2}$\, (GeV) & 1037 & 1249 & 1149 & 1107 & 1219 & 994\\
$\tan \beta$ & 2 & 2 & 3 & 10 & 3 & 2\\
$\mu$\, (GeV) & 158.2 & 160.4 & 165.8 & 174.1 & 159.5 & 201.6\\
$v_S$\, (GeV) & -19.0 & -12.7 & -8.4 & -0.3 & -5.0 & -13.7\\
$v_T$\, (GeV) & 1.5 & 1.5 & 1.1 & 0.3 & 1.9 & 1.1\\ \hline
$m_{h_1}$\, (GeV) & 126.6 & 122.8 & 123.2 & 123.6 & 127.4 & 122.2\\
$m_{h_2}$\, (GeV) & 457.7 & 312.3 & 308.5 & 421.8 & 465.8 & 487.5\\
$m_{h_3}$\, (GeV) & 1282.2 & 799.7 & 935.2 & 1251.6 & 1176.5 & 762.3\\
$m_{h_4}$\, (GeV) & 2667.0 & 2965.0 & 3350.8 & 7104.7 & 2847.1 & 2933.0\\
$m_{H_1^\pm}$\, (GeV) & 1284.0 & 793.5 & 933.4 & 1257.7 & 1174.3 & 765.4\\
$m_{\tilde{\chi}_1^0}$\, (GeV) & 106.3 & 110.4 & 120.5 & 172.6 & 128.0 & 134.5\\
$m_{\tilde{\chi}_1^\pm}$\, (GeV) & 168.4 & 169.6 & 177.3 & 193.1 & 173.1 & 211.5\\
$m_{W}$\, (GeV) & 80.363 & 80.365 & 80.362 & 80.361 & 80.369 & 80.362\\\hline\hline
\end{tabular}
\caption{\label{TAB:AMDGSSM} Benchmark points with a light singlet in the Aligned MDGSSM}.
\end{table}

\subsection{W mass in the aligned DGNMSSM}
\label{sec:WADGNSSM}

We turn now to the case of the aligned DGNMSSM described in section \ref{sec:scenarios}. We set the couplings $\lambda_S, \lambda_T$ to their $N=2$ values and then choose $m_{DY}$ and $T_S$ to make $\Delta_{hs}$ and $\Delta_{Hs}$ vanish. This leads to
\begin{align}
  m_{DY} =& c_{2\beta} \mueff \nn\\
  T_S =& -g_Y (\kappa v_S - \sqrt{2} s_{2\beta} \mueff).
\end{align}
This has an interesting consequence because in this model
\begin{align}
  \Bmueff =& \frac{1}{\sqrt{2}} T_S v_S + \frac{1}{2} \lambda_S \kappa v_S^2 
  = - \frac{1}{2\sqrt{2}} g_Y\kappa v_S^2 + g_Y s_{2\beta} v_S  \mueff.
\end{align}
Then
\begin{align}
  M_A^2 =& g_Y v_S (2 \mueff - \frac{\kappa v_S}{\sqrt{2}s_{2\beta}} ) = g_Y v_S^2 (g_Y - \frac{\kappa}{\sqrt{2}s_{2\beta}} ).
\end{align}

\begin{figure}\centering
  \includegraphics[width=0.45\textwidth]{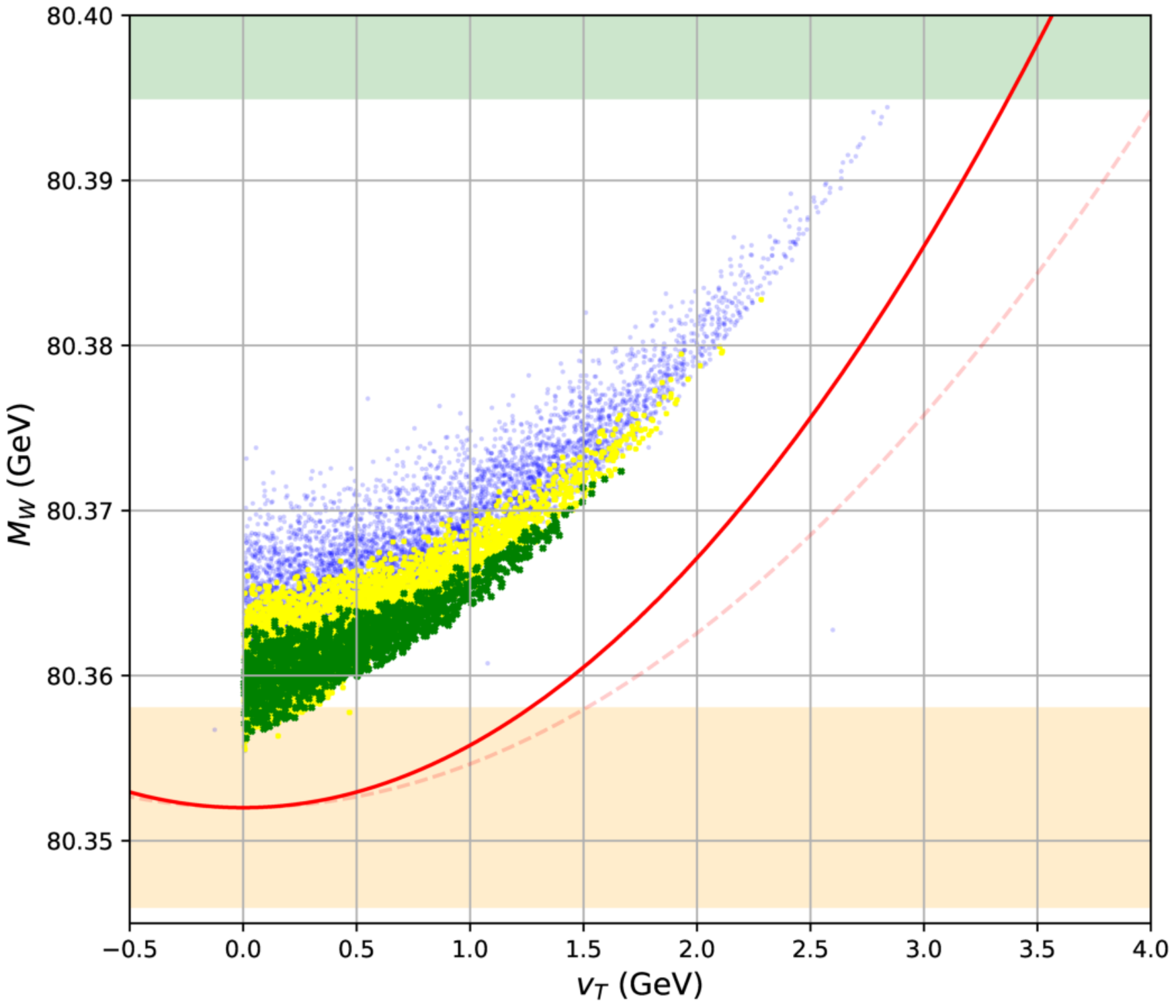} \includegraphics[width=0.45\textwidth]{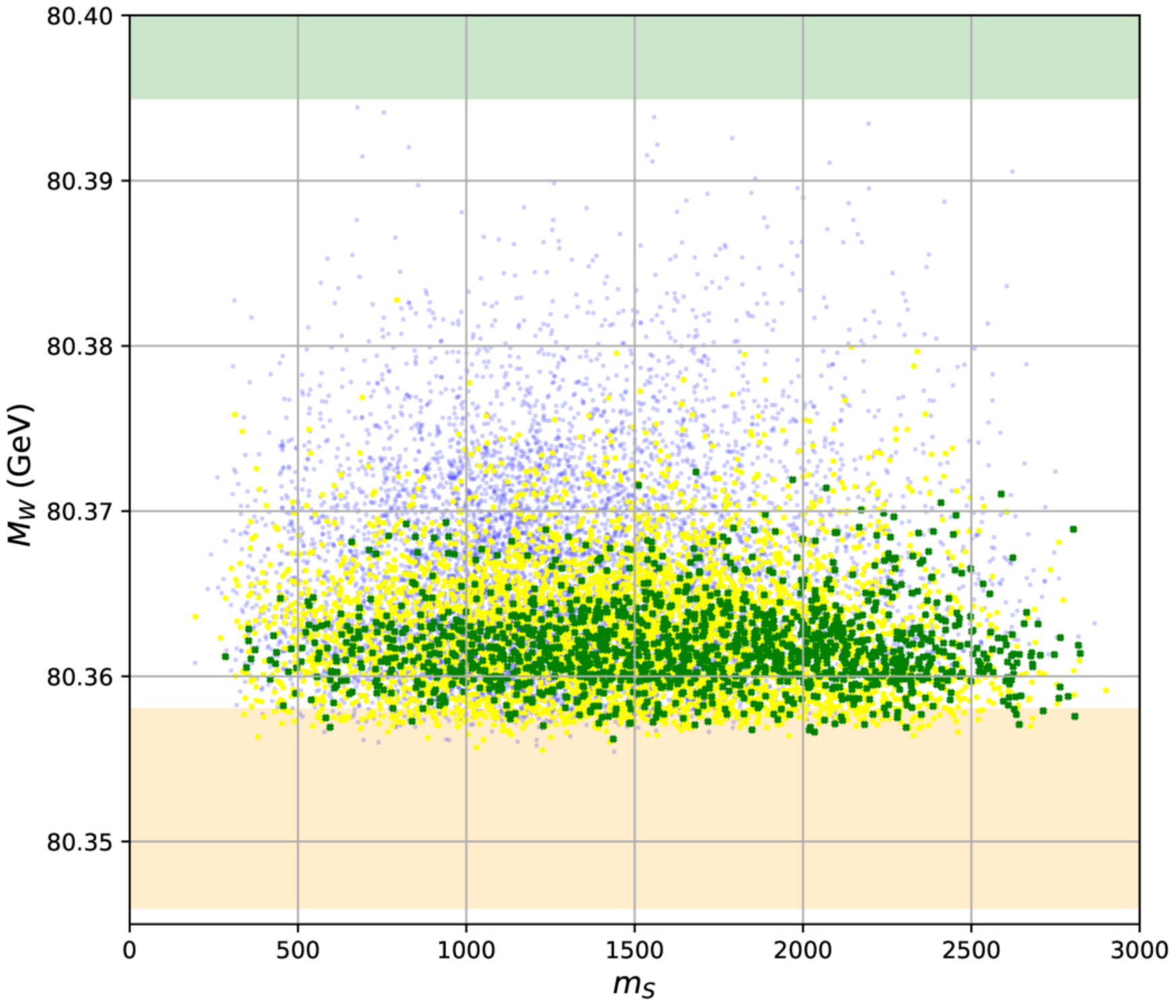}
 \caption{\label{FIG:ADGNMSSM} W boson mass in the aligned DGNMSSM. Left: W boson mass against triplet expectation value. Right: W boson mass against singlet-like Higgs mass. Colour coding of points is described in the text.}
\end{figure}

We perform a scan using the same strategy as the previous sections; we use a common mass for the third generation squarks $M_{\rm SUSY};$ we set $m_{Q, 33}^2 = m_{U,33}^2 = m_{D,33}^2 = M_{\rm SUSY}^2$ (the other squarks and sleptons we retain fixed at $\sqrt{10}$ TeV). Then we scan with the parameter ranges, using the same likelihood function as before:
\begin{align}
  M_{\rm SUSY} \in [2000,10000]&\, \GeV, \nn\\
  \quad m_{D2} \in [400,1500]&\, \GeV, \quad v_S \in [-1500, 1500]\, \GeV, \quad v_T \in [-5, 5]\, \GeV \nn\\
  \quad \kappa \in [-1.5,1.5],& \quad T_\kappa \in [-2000, 2000]\, \GeV, \quad \tan \beta \in [2,50].
\end{align}
In our scans we impose that all Higgs searches are satisfied using {\tt HiggsBounds} and {\tt HiggsSignals}. We show the results for the W boson mass in figure \ref{FIG:ADGNMSSM} where the points have the same colour coding as in the previous sections. It is apparent that in this model it is complicated to enhance the W boson mass. This is because we have only a small quantum effect from $\lambda_S, \lambda_T$, but also because the lightest neutralinos are typically rather light: since $\mueff = g_Y v_S/2$ we need a large $v_S \gtrsim 500$ GeV (or $1$ TeV for our more stringent points) to have heavy enough higgsinos. Then we need $\kappa$ negative and not too small to avoid a too-small pseudoscalar/charged Higgs mass (if we neglect $\kappa$ then $M_A$ is bounded by $g_Y v_S$, so $M_A > 600$ GeV requires $v_S \gtrsim 1700$ GeV, at the limit of our search range). So this implies that the singlino is generally heavy compared to $m_{DY}$:
\begin{align}
\frac{m_{DY}}{\sqrt{2} \kappa v_S} =& \frac{g_Yc_{2\beta}}{2 \sqrt{2} \kappa} .
\end{align}

In figure \ref{FIG:ADGNMSSM} we show the scan results with the same colour coding as in previous sections. It is clear that in this scenario, models which can explain a large $W$ mass are driven by a larger $v_T$ with some modest quantum corrections enhancing the mass by $\mathcal{O}(10)$ MeV; there is very little spread due to the lack of variation in $\lambda_T$. However, it is difficult to find points with large enough $v_T$ that satisfy other bounds.

Another feature of the selected points is that in almost all cases $v_T >0$; considering $T_T = M_T =0$ in equation (\ref{minvT}) means that in order for $\tilde{m}_{TR}^2 >0$ (so that, at least, the pseudoscalar triplet should be non-tachyonic, since we take $B_T=0$) and $v_T < 0$ we would need $\mueff > |m_{D2} c_{2\beta}|$. But we need large $\tan \beta$ to obtain the correct Higgs mass, and therefore $g_Y v_S/2 \gtrsim m_{D2}$; for our selected points we require a minimum of $m_{D2} > 600$ GeV, and so $v_S$ would again be beyond our search range.  

Since we are interested here in alignment, we may have a light singlet scalar without falling foul of either light Higgs or heavy Higgs searches. In this limit we have
\begin{align}
\tilde{m}_S^2 \rightarrow& \frac{1}{4} \bigg[ 2 v_S (\sqrt{2} T_\kappa + 3 \sqrt{2} \kappa M_S + 4 v_S \kappa^2) +v^2 g_Y(g_Y - \sqrt{2} \kappa s_{2\beta})\bigg].      
\end{align}
$T_\kappa$ of opposite sign to $v_S$ allows the singlet to be made light while making $M_A$ arbitrarily heavy. In the scan, we do not impose any likelihood bias to search for points with a light singlet, but we show benchmark points passing all constraints which have light singlet masses in table \ref{BenchmarksDGAN}. They show a W mass consistent with the SM prediction.

\begin{table}\centering
\begin{tabular}{|c|c|c|c|c|} \hline \hline
  & A-DGN1 & A-DGN2 & A-DGN3 & A-DGN4 \\ \hline \hline
$M_{\rm SUSY}$\, (GeV) & 6368.4 & 5186.5 & 8219.5 & 8702.9\\
$m_{D2}$\, (GeV) & 802 & 991 & 932 & 923\\
$\tan \beta$ & 26 & 11 & 29 & 22\\
$\kappa$ & -0.418 & -0.332 & -0.446 & -0.418\\
$v_S$\, (GeV) & 1417.4 & 1402.4 & 1337.0 & 1488.8\\
$v_T$\, (GeV) & 0.2 & 0.1 & 0.1 & 0.8\\
$T_\kappa$\, (GeV) & -595.6 & -267.5 & -648.0 & -639.4\\ \hline\hline
$m_{h_1}$\, (GeV) & 124.1 & 122.9 & 125.2 & 123.4\\
$m_{h_2}$\, (GeV) & 353.5 & 417.6 & 347.2 & 353.9\\
$m_{h_3}$\, (GeV) & 1838.5 & 1284.5 & 1843.0 & 1803.0\\
$m_{h_4}$\, (GeV) & 6710.7 & 12226.0 & 10817.9 & 4056.2\\
$m_{H_1^\pm}$\, (GeV) & 1841.4 & 1288.5 & 1846.1 & 1806.2\\
$m_{\tilde{\chi}_1^0}$\, (GeV) & 72.2 & 84.2 & 64.5 & 75.2\\
$m_{\tilde{\chi}_1^\pm}$\, (GeV) & 280.1 & 274.6 & 265.1 & 291.9\\
$m_{W}$\, (GeV) & 80.362 & 80.361 & 80.361 & 80.363\\\hline\hline
\end{tabular}
\caption{\label{BenchmarksDGAN} Benchmark points for the ``aligned DGNMSSM''. Input parameters are given above the double line, and masses of the most important particles below. }
\end{table}

\subsection{W mass in the general DGNMSSM}
\label{sec:WDGNSSM}

Finally we consider the general DGNMSSM, where we allow the values of $\lambda_S, \lambda_T$ to vary and do not fix the values of $m_{DY}$ or $T_S$ to require alignment but scan over them. This means that we will not focus on light singlet (or doublet) scalars. Similar to the aligned case, there is still a see-saw effect on the lightest neutralino mass due to the non-zero singlino mass, which can drive down the quantum corrections to the W boson, but a large $|\lambda_T|$ can compensate for this and also help enhance the SM-like Higgs mass.


\begin{figure}\centering
  \includegraphics[width=0.45\textwidth]{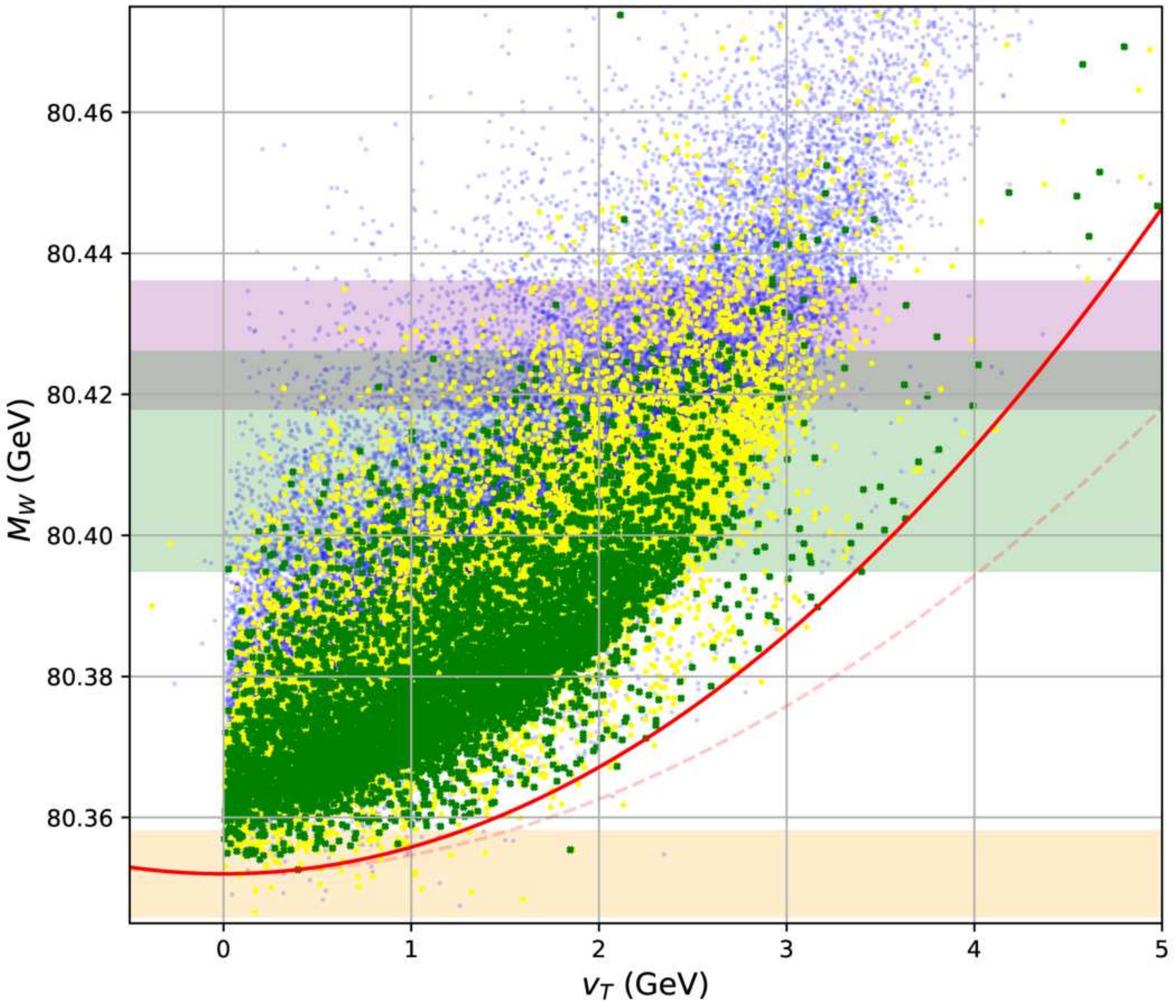} \includegraphics[width=0.45\textwidth]{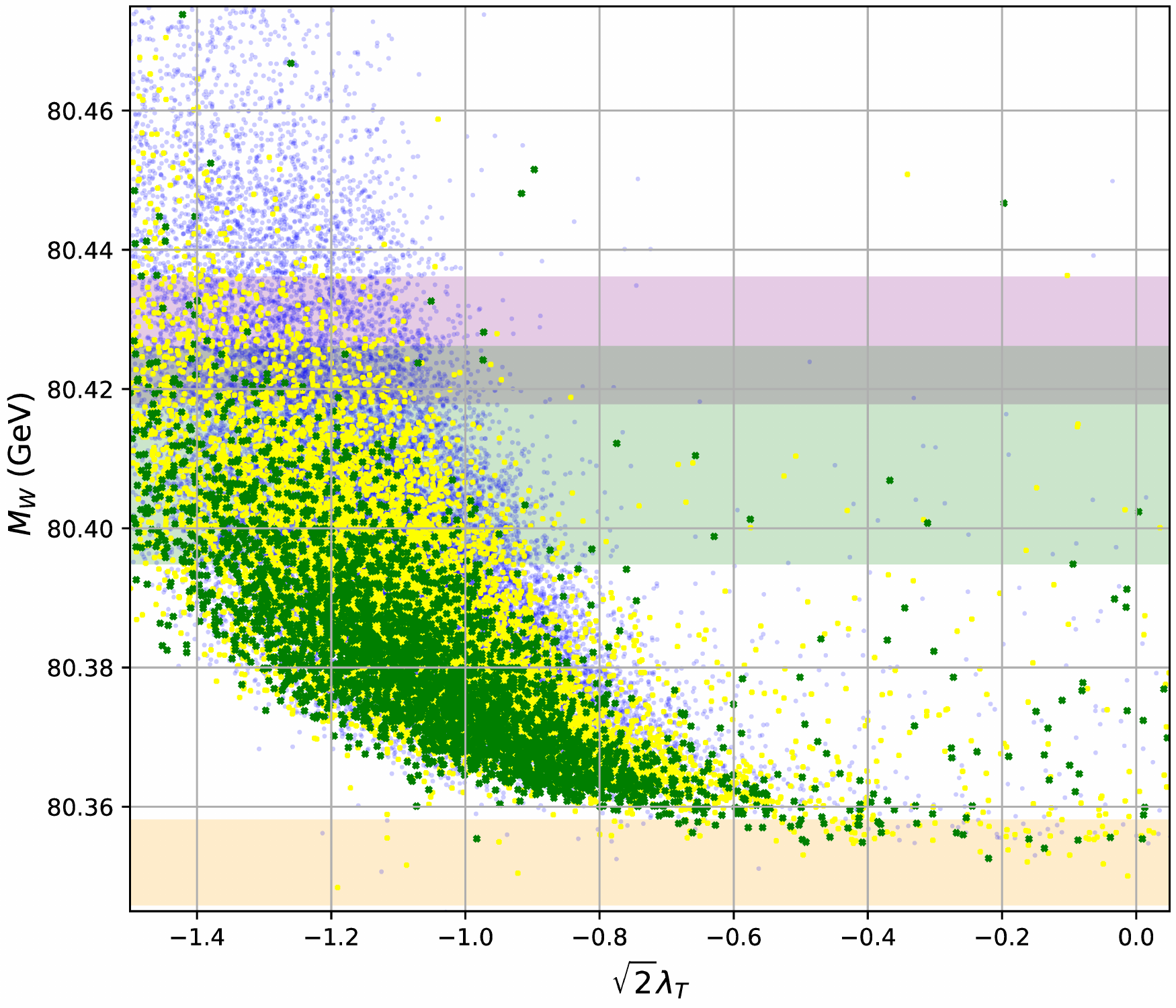}
 \caption{\label{FIG:DGNMSSM} W boson mass in the general DGNMSSM. Left: W boson mass against triplet expectation value. Right: W boson mass against $\lambda_T$. Colour coding of points as in previous figures.}
\end{figure}

\begin{table}\centering
\begin{tabular}{|c|c|c|c|c|c|}\hline \hline
  & DGN1 & DGN2 & DGN3 & DGN4 & DGN5 \\ \hline \hline
$m_{DY}$\, (GeV) & 392 & 298 & 410 & 380 & 292\\
$m_{D2}$\, (GeV) & 927 & 971 & 841 & 1003 & 805\\
$\kappa$ & 1.391 & -1.369 & -1.266 & -1.309 & -1.361\\
$\tan \beta$ & 9 & 23 & 21 & 30 & 30\\
$-\lambda_S$ & 0.727 & -0.893 & -0.544 & 0.554 & -0.677\\
$\sqrt{2}\lambda_T$ & -1.426 & 1.496 & 1.463 & -1.303 & -1.296\\
$-T_S$\, $(\GeV)$ & 3077 & 3747 & -2496 & -3002 & 1183\\
$T_\kappa$\, $(\GeV)$ & 1139 & 350 & -1292 & -571 & 728\\
$v_S$\, (GeV) & -658.1 & -539.1 & 574.5 & 524.7 & -482.8\\
$v_T$\, (GeV) & 2.7 & 2.3 & 1.7 & 2.5 & 2.2\\
$m_{h_1}$\, (GeV) & 125.1 & 125.3 & 125.8 & 124.6 & 124.7\\
$m_{h_2}$\, (GeV) & 1017.6 & 937.1 & 665.0 & 831.8 & 740.1\\
$m_{A_1}$\, (GeV) & 757.6 & 93.7 & 778.1 & 115.1 & 502.2\\
$m_{H_1^\pm}$\, (GeV) & 2793.1 & 3195.1 & 1281.7 & 778.5 & 806.5\\
$m_{\tilde{\chi}_1^0}$\, (GeV) & 115.0 & 87.2 & 123.6 & 110.9 & 90.7\\
$m_{\tilde{\chi}_1^\pm}$\, (GeV) & 278.8 & 273.6 & 265.5 & 254.8 & 268.8\\
$m_{W}$\, (GeV) & 80.421 & 80.421 & 80.424 & 80.420 & 80.422\\\hline\hline
\end{tabular}
\caption{\label{BenchmarksDGN} Benchmark points for the general DGNSSM. Input parameters are given above the double line, and masses of the most important particles below. }
\end{table}

We perform a scan using the strategy as in sections \ref{sec:MSSMnomu} and \ref{sec:WMDGSSM}, with the parameter ranges:
\begin{align}
  m_{DY} \in [100,700]\, \GeV,& \quad m_{D2} \in [150,1200]\, \GeV, \quad v_S \in [-700, 700]\, \GeV, \quad v_T \in [-5, 5]\, \GeV \nn\\
  \qquad \kappa \in [-1.5,1.5], \qquad& T_\kappa \in [-2000, 2000]\, \GeV, \qquad T_S \in [-4000, 4000]\, \GeV, \nn\\
  \qquad \lambda_S \in [-1.5,1.5],& \qquad \sqrt{2}\lambda_T \in [-1.5,1.5], \qquad \tan \beta \in [2,50].
\end{align}
We give plots in figure \ref{FIG:DGNMSSM} with the same colour coding as in the previous sections; the difference in the distribution to the previous examples is rather striking. It is clear that in this scenario a large negative $\lambda_T$ and a positive $v_T$ is favoured; this gives a tree-level enhancement to the Higgs mass and a loop-level enhancement to the $W$-boson mass. The singlino component will mix less with the higgsinos than in the MDGSSM because of the $\sqrt{2} \kappa v_S$ singlino mass, and thus the effect of $\lambda_S$ on the $W$ mass is diminished. The asymmetry in the signs of $\lambda_T$ and $v_T$  can be explained by the fact that we only take positive Dirac gaugino masses in the scans.


\section{Conclusions}
\label{SEC:CONCLUSIONS}
%
%
%

We have shown that an aligned Dirac Gaugino NMSSM is possible and compatible with current collider constraints; it can even lead to relatively light singlet scalars that may be of interest to future searches (although would be rather difficult to find directly as they are difficult to produce). Such a model favours a W boson mass compatible with or just above the SM prediction. We also showed how two different Dirac Gaugino scenarios can easily be compatible with an enhanced W boson mass, including a precise computation of the quantum corrections for the first time, which are now incorporated automatically in the package \SARAH. We also used this computation to add more nails to the coffin of the ``MSSM without $\mu$ term.''

We have been conservative in our application of collider constraints and concluded that the MDGSSM models would typically contain underdense dark matter densities. However, it would be interesting to examine the issue of dark matter and collider constraints again in all of these classes of models when all the latest searches for electroweakinos have been recast; we have provided ample benchmark points for this purpose. In the DGNMSSM or its aligned version, if we impose strict R-parity or have a heavy gravitino (by no means entirely obvious assumptions), it may be that we require a Higgs funnel to obtain the correct relic density, which would require a sophisticated search strategy to find allowed parameter ranges, along e.g. the lines of \cite{Goodsell:2022beo}. However, it is also likely that a light singlino in the aligned DGNMSSM could fulfil the role of the Higgs funnel. We leave these questions to future work.

%
%

%
\section*{Acknowledgments}
%
%

The work of W.~K. is supported by the Contrat Doctoral Sp\'ecifique Normalien (CDSN) of Ecole Normale Sup\'erieure -- PSL. K.~B., P.~S. and M.~D.~G. acknowledge support from the grant
\mbox{``HiggsAutomator''} of the Agence Nationale de la Recherche
(ANR) (ANR-15-CE31-0002). M.~D.~G. also acknowledges support from the ANR via grant \mbox{``DMwithLLPatLHC''}, (ANR-21-CE31-0013). M.~D.~G. thanks Werner Porod and Hannah Day for interesting discussions. W.~K. thanks Pierre Fayet for illuminating discussions during the early stages of this work.

\bibliographystyle{utphys}
\bibliography{KMPW}

\end{document}